\documentclass[conference]{IEEEtran}
\IEEEoverridecommandlockouts
\usepackage[usenames,dvipsnames,svgnames,x11names]{xcolor}

\usepackage{amsmath,amssymb,amsfonts}
\usepackage{algorithmic}
\usepackage{graphicx}
\usepackage{caption}
\usepackage{subcaption}
\usepackage{textcomp}
\usepackage{braket}

\usepackage{tikz}
\usetikzlibrary{quantikz}
\usepackage{adjustbox}

\def\BibTeX{{\rm B\kern-.05em{\sc i\kern-.025em b}\kern-.08em -
    T\kern-.1667em\lower.7ex\hbox{E}\kern-.125emX}}
\begin{document}

\title{Parallelizing Quantum-Classical Workloads: Profiling the Impact of Splitting Techniques}

\author{
\IEEEauthorblockN{
Tuhin Khare\IEEEauthorrefmark{1}
Ritajit Majumdar\IEEEauthorrefmark{2}, 
Rajiv Sangle\IEEEauthorrefmark{1},
Anupama Ray\IEEEauthorrefmark{2}, \\
Padmanabha Venkatagiri Seshadri\IEEEauthorrefmark{8}, and
Yogesh Simmhan\IEEEauthorrefmark{1}
}

\IEEEauthorblockA{\IEEEauthorrefmark{1} Indian Institute of Science (IISc), Bengaluru}
\IEEEauthorblockA{\IEEEauthorrefmark{2} \textit{IBM Quantum}, IBM India Research Lab}
\IEEEauthorblockA{\IEEEauthorrefmark{8} IBM India Research Lab}
\IEEEauthorblockA{Email: tuhinkhare@iisc.ac.in,  majumdar.ritajit@ibm.com, rajivsangle@iisc.ac.in}
\IEEEauthorblockA{ anupamar@in.ibm.com,  seshapad@in.ibm.com,  simmhan@iisc.ac.in}

}

\maketitle

\begin{abstract}
Quantum computers are the next evolution of computing hardware.
Quantum devices are being exposed through the same familiar cloud platforms used for classical computers, and enabling seamless execution of hybrid applications that combine quantum and classical components. Quantum devices vary in features, e.g., number of qubits, quantum volume, CLOPS, noise profile, queuing delays and resource cost. So, it may be useful to split hybrid workloads with either large quantum circuits or large number of quantum circuits, into smaller units. In this paper, we profile two workload splitting techniques on IBM's Quantum Cloud: (1) \textit{Circuit parallelization}, to split one large circuit into multiple smaller ones, and (2) \textit{Data parallelization} to split a large number of circuits run on one hardware to smaller batches of circuits run on different hardware. 
These can improve the utilization of heterogenous quantum hardware, but involve trade-offs. 
We evaluate these techniques on two key algorithmic classes: Variational Quantum Eigensolver (VQE) and Quantum Support Vector Machine (QSVM), and measure the impact on circuit execution times, pre- and post-processing overhead, and quality of the result relative to a baseline without parallelization. Results are obtained on real hardware
and complemented by simulations.  
We see that (1) VQE with circuit cutting is $\sim 39\%$ better in ground state estimation than the uncut version, and (2) QSVM that combines data parallelization with reduced feature set yields upto $3\times$ improvement in quantum workload execution time and reduces quantum resource use 
by $3\times$, while providing comparable accuracy.
Error mitigation can improve the accuracy by $\approx 7\%$ and resource foot-print by $\approx 4\%$ compared to the best case among the considered scenarios.  

\end{abstract}

\section{Introduction}
\label{sec:intro}

The long-held promise of quantum computing is starting to come true with the launch of real quantum hardware for remote access by leading cloud service providers (CSP) -- Amazon AWS \cite{aws_quantum}, Google Cloud \cite{google_quantum}, IBM \cite{ibm_quantum} and Microsoft Azure \cite{azure_quantum}.
In practice, quantum applications use the quantum hardware only for specialized kernels or routines which they are designed for and the rest of the application runs on ``classical'' (binary, general-purpose) computers. Such a hybrid execution model is well-suited for a cloud-based deployment where both classical and quantum resources can be co-located.
%

Adoption of this emerging hybrid quantum-classical computing paradigm requires a wide base of users to develop the necessary algorithmic and programming skills, which will benefit future enterprise workforce. It can also open up opportunities for academic researchers to explore novel problems.
In that regard, many CSPs are providing access to their quantum hardware through both limited free offerings of low-capacity quantum machines, and time-limited reservations through subscriptions or a pay-as-you-go model for higher-capacity quantum machines. 
Just as classical computers are heterogeneous (x86, x64, ARM, etc.), quantum devices also differ in the number of qubits, quantum volume and CLOPS, and their noise profile (see Table~\ref{tbl:machineprofile}). This diversity extends to their cost for use and availability, as observed through queuing delays to these scarce resources~\cite{ccgrid2023showcase}.

\begin{table}[htb]
\caption{Quantum hardware machine profiles}
\label{tbl:machineprofile}
\begin{center}
\begin{tabular} {|p{1.55cm}||p{0.7cm}|p{1.7cm}|p{0.9cm}|p{1.7cm}|}
\hline
\bf HW Name & \bf Qubits & \bf Quantum Vol.& \bf CLOPS & \bf Proc. Type \\
\hline
ibm\_hanoi & 27 & 64 & 2.3k & Falcon r5.11 \\
\hline
ibmq\_jakarta & 7 & 16 & 2.4k & Falcon r5.11H \\
\hline
 ibm\_oslo & 7 & 32 & 2.6k & Falcon r5.11H \\
 \hline
ibm\_nairobi & 7 & 32 & 2.6k & Falcon r5.11H \\
 \hline
 ibmq\_perth & 7 & 32 & 2.9k & Falcon r5.11H \\
\hline
ibmq\_manila & 5 & 32 & 2.8k & 	Falcon r5.11L \\
\hline
ibmq\_quito & 5 & 16 & 2.5k & Falcon r4T \\
\hline
ibmq\_belem & 5 & 16 & 2.5k & Falcon r4T \\
\hline
ibmq\_lima & 5 & 8 & 2.7k & Falcon r4T \\
\hline
\end{tabular}
\end{center}
\end{table}

Under budget constraints, a user has to make the best use of available resources, which could be either: (1) several low-capacity machines, or (2) a mix of time-limited access to high-capacity machines complemented by free low-capacity machines. Such users can benefit from workload splitting techniques to break their workload into smaller parts. This brings to fore the following question: \emph{What are the overheads in execution time, queuing delay and quality of result incurred from workload splitting techniques?}

There is a need to explore these overheads from an application point-of-view. We address this need and by considering two prominent classes of hybrid algorithms: First is from the class of Variational Quantum Algorithms (VQAs) known as \textit{Variational Quantum Eigensolver (VQE)}~\cite{cerezo2021variational}. VQE is extensively used for ground-state energy estimation of molecules, and the size of the circuit usually scales linearly with the size of the system. The second is a Quantum Machine Learning (QML) algorithm, \textit{Quantum Support Vector Machine (QSVM)}, that 
depends on data for its training and testing. The number of circuits to train QSVM is $\mathcal{O}(n^2)$, given $n$ training data points. Thus, even for $100$ data points, a quantum device needs to run $10,000$s of circuits, with classical optimization done in each iteration. Also, quantum algorithms are probabilistic, and need to be executed multiple times to obtain a reliable expectation value of the result. All of these together is computationally demanding, given the current limitations of quantum hardware \cite{preskill2018quantum}.

In our study, we apply two hardware-independent splitting techniques to the above use-cases, namely: (1) \textit{circuit parallelization} \cite{peng2020simulating} that allows a bigger circuit to be partitioned into multiple smaller ones, each of which can be computed independently on separate hardware, and (2) \textit{data parallelization}, where the same circuit is run on different parameters on multiple diverse quantum hardware (\S~\ref{sec:workfload-splitting}). The focus of the study is to conduct detailed experiments on real quantum hardware to \textit{profile the performance and quality impact of these techniques}, relative to a baseline which does not use splitting (\S~\ref{sec:evaluation}). Specifically, for circuit parallelization, we compare the performance of the platform, workload metrics and quality of result, post-convergence, of the circuit-cut VQE run on multiple machines against the baseline uncut-circuit run on one machine. For data parallelization, we partition circuits generated for the QSVM kernel into batches and run them on quantum machines with diverse capacities, and compare their performance against a baseline running on one machine. 
Our results of profiling indicate these key observations:
\begin{enumerate}
    \item VQE with circuit cutting and parallelization achieves a $\approx 39\%$ improvement in ground state estimation than the uncut version.
    \item QSVM with using both data parallalization and reduced feature set yields up to $3\times$ faster quantum workload execution time and reduces quantum resource usage (qubits-seconds) by $3\times$, while providing comparable accuracy with respect to the baseline. Error mitigation could improve the accuracy to nearly 7\% of the best case (among the considered scenarios) while improving resource foot-print by 4\% compared to best case.
    \item Classical overheads vary depending on the application. While it is on the order of 100~ms for QSVM, it goes up to 30~mins for VQE. 
\end{enumerate}

\section{Related Work}
\label{sec:related}

In the classical computing world, there is an extensive body of work in scheduling workflows on HPC clusters and cloud resources~\cite{adhikari2019survey}. These typically attempt to schedule the components tasks of a Directed Acyclic Graph (DAG) onto different compute resources, taking into account the execution time of tasks on heterogeneous resources~\cite{topcuoglu2002performance}, data and control dependencies, data movement costs, monetary cost~\cite{yu2005cost} and even energy~\cite{bousselmi2016energy}. However, these approaches cannot be applied directly to quantum cloud platforms.

Recent profiling studies~\cite{ravi2021} have investigated long term traces of generic workloads to understand quantum platform characteristics such as queuing delay, calibration cycles and error profiles. However these studies predate practical quantum workload parallelization approaches like circuit splitting~\cite{peng2020simulating}. Further, the study does not capture the use-case specific impact of parallelization techniques on performance. In \cite{zhang2022, parekh2021} the authors propose a resource scheduling approach to map users tasks to resources. But it does not focus on any form of parallelization based profiling.

Circuit knitting is an umbrella term used to encompass various techniques to split a large quantum circuit into multiple smaller ones. Two broad subdomains of this are: (i) cutting the wire of a circuit to create multiple smaller fragments (termed as \textit{circuit cutting} henceforth)~\cite{peng2020simulating, perlin2021quantum, tang2021cutqc}
and (ii) replacing two qubit gates by a local single qubit gate and classical feedforward communication~\cite{brenner2023optimal, piveteau2022circuit, marshall2022}. Both of these methods usually lead to smaller fragments (or subcircuits) which can be independently executed on smaller hardware. However, to the best of our knowledge, there are no studies which focus the performance impact of the parallelization techniques on specific application classes. In this paper, we restrict ourselves to only circuit cutting.

Efforts have been made previously to distribute the computing workload both in quantum and classical computing. Ensembled Quantum Computing (EQC)~\cite{eqc} is an approach where the quantum workload is distributed across multiple machines. The purpose of EQC is to offer robustness against both temporal variations in noise and also across non-homogeneity of quantum hardware. However, it focuses only on VQA class of algorithms and does not explore the impact of various parameters of the workload, such as circuit cutting, feature space reduction to reduce qubit requirements, shots required for convergence, and the need for error mitigation.

\section{Background}
\label{sec:background}

Current quantum devices are noisy, and can accommodate only low-depth circuits. Therefore, hybrid quantum-classical algorithms have been developed to split the overall computing workload into a quantum and a classical processing unit, with back-and-forth interactions between these two resources. These algorithms, in general, conform to parameterized quantum circuits where the parameters are initialized randomly, and are updated in each iteration by a classical optimizer to converge towards the optimal expectation value.

IBM Qiskit Runtime~\cite{QiskitRuntime} model has been proposed for co-location of classical resources with the quantum hardware. This allows classical primitives which co-ordinate with quantum circuits to iterate with low-latency and efficiently. Moreover, Qiskit Runtime also provides inbuilt error suppression~\cite{viola1998dynamical} and mitigation methods (e.g., Measurement error mitigation~\cite{van2022}, Zero Noise Extrapolation~\cite{temme2017} or Probabilistic Error Cancellation~\cite{van2022probabilistic}), thus allowing more faithful circuit execution on noisy hardware. 

We leverage this support for hybrid quantum-classical workloads to profile the VQE and QSVM use-cases. The two use-cases are briefly described below.

\textbf{Hamiltonian simulation using VQE.~}
VQE is a widely studied hybrid quantum-classical algorithm to evaluate the minimum expectation value of some Hamiltonian $H$. A VQE is characterized by a parameterized circuit (or ansatz) $\ket{\psi(\Vec{\theta})}$, where $\Vec{\theta} = \{\theta_1, \theta_2, \hdots, \theta_m\}$ denotes the set of parameters. These parameters are initialized either based on some domain knowledge or randomly, and then updated on each iteration by a classical optimizer to minimize the energy of the system given by the expectation value $\braket{\psi(\Vec{\theta})|H|\psi(\Vec{\theta})}$ \cite{peruzzo2014variational}.

In this study, we consider a simple $n$ qubit Quantum Heisenberg Spin Model (QHSM) where the interaction is between neighbouring qubits along the $Z$-direction with coupling strength $J_z = 1$. The Hamiltonian of such a model is:

\begin{equation}
\label{eq:qhsm_zz}
    H = \sum_{i=1}^{n-1} \sigma_{i}^{z}\sigma_{i+1}^{z}
\end{equation}
For the rest of the paper, we shall use $\sigma^z$ and $Z$ interchangeably to denote the Pauli-Z operator.

\textbf{Classification using QSVM. }
Havlicek et al.~\cite{QSVM} propose two ways of creating a quantum version of the 
classical Support Vector Machine (SVM) called the Quantum SVM (QSVM). One is by leveraging variational quantum circuits to train a variational quantum SVM. The second approach, which we use in this paper,
is called \textit{Kernel Estimation} algorithm, and it estimates the kernel function and optimizes the classifier directly.

In classical SVM, feature maps are used to perform non-linear transformations to a higher dimension if the training data are unable to be separated in a lower dimension. In this feature space, computing the distance between data points for classification is equivalent to computing the inner product of each pair of data points.
This collection of inner products is called a \textit{kernel}. In QSVM, the aim is to compute a kernel that was hard to compute classically, and 
take advantage of the large dimensionality of the quantum Hilbert space for enhanced classification. We couple this quantum kernel with convex optimization on a classical computer, thus making this a hybrid application. 

A key challenge in training QML algorithms on quantum devices is the inability to process a lot of data. Unlike classical ML algorithms which are trained on millions of data points, training a QSVM requires us to process $n^2$ circuits for $n$ data points, while testing takes $n \cdot m$ circuits, $m$ being the number of test data points. Thus, the quantum computer is used twice here -- once for computing the training kernel and once for the testing kernel, while the rest of the steps are all classical. The following steps illustrate the workflow when a QSVM code is submitted to a hybrid quantum-classical system:

\begin{itemize}
\item Pre-process the data on classical hardware.
\item Form circuit batches with $n^2$ circuits (to create training Kernel) and $n \cdot m$ circuits (to create testing Kernel), and submit the jobs (batches of circuits).
\item Job is queued on quantum cloud, circuits are validated, quantum provider transpiles the circuits in a job and then run the circuits.
\item Optimizer is a Classical SVM formulated in terms of a dual quadratic program that needs only access to the kernels. This optimizer finds an optimal $\alpha$ to construct the classifier and classification is then performed using classical SVM as optimizer.

\end{itemize}

Given the large number of circuits, 
training on a real quantum hardware is time-consuming. Current hardware divides the total number of circuits to run into batches called jobs, 
and runs these jobs serially. 
So, the program needs to wait for all circuits to complete execution before it can construct the kernel. 
Also practical engineering challenges restricts execution of a large number of circuits. The code waits for these jobs to wait on a queue (which can vary significantly based on the load/status of a particular hardware) 
and then get executed. A job can time-out if the wait period is too long, causing the application to fail.

\section{Workload Splitting Techniques}
\label{sec:workfload-splitting}
In this study we consider two methods for workload splitting: (i) partitioning a circuit itself into smaller fragments, and (ii) data/circuit parallelization where we split a collection of circuits into multiple batches (which can be run in parallel), with multiple circuits per batch. 

\subsection{Circuit Cutting}
\label{sec:workfload-splitting:ckt-cut}

Circuit cutting is a method to partition a quantum circuit into multiple smaller fragments, each of which can be executed independently and parallelly on a smaller hardware~\cite{peng2020simulating, tang2021cutqc}. Given a circuit (or a quantum channel) $\Phi$, the expectation value of some observable $A$ can be evaluated as $\Phi(A) = \sum_i c_i \Phi_i(A)$, where $\Phi_i(A) = Tr\{AO_i\}\rho_i$, $O_i$ and $\rho_i$ being tomographically complete measurement and preparation basis respectively, and $c_i \in \{\frac{1}{2}, -\frac{1}{2}\}$~\cite{peng2020simulating}. Each fragment $\Phi_i$, in general, contains fewer qubits and gates than $\Phi$ and is expected to (i) have lower transpilation time, (ii) eliminate a few SWAP gates by better transpilation of the circuit \cite{basu2021qer}, and (iii) lower the noise in the system in certain scenarios \cite{ayral2021quantum, basu2021qer, majumdar2022error}. 

In \cite{tang2021cutqc}, the authors showed that the sets $O_i \in \{X,Y,Z\}$ and $\rho_i \in \{\ket{0},\ket{1},\ket{+},\ket{+i}\}$ are tomographically complete, and hence are sufficient for the circuit cutting purpose. The overall expectation value $\Phi(A)$ is calculated using post-processing over the fragments using classical computers. However, a bottleneck of this method is that the classical post-processing time scales exponentially with the number of cuts, and hence this method is feasible only for circuits which can be partitioned using a small number of cuts.

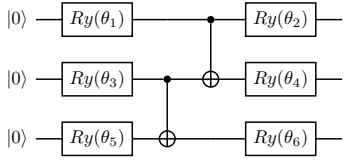
\begin{figure}[htb]
    \centering
    \begin{adjustbox}{scale=0.7}
    \begin{quantikz}
    \lstick{$\ket{0}$}& \gate{Ry(\theta_1)} & \qw & \ctrl{1} & \gate{Ry(\theta_2)} & \qw \\
    \lstick{$\ket{0}$}& \gate{Ry(\theta_3)} & \ctrl{1} & \targ{} & \gate{Ry(\theta_4)} & \qw \\
    \lstick{$\ket{0}$}& \gate{Ry(\theta_5)} & \targ{} & \qw & \gate{Ry(\theta_6)} & \qw
    \end{quantikz}
    \end{adjustbox}
    \caption{A template of a 3-qubit RealAmplitudes ansatz with single repetition and reverse-linear entanglement}
    \label{fig:realamplitude}
\end{figure}

Circuit cutting has previously been used on toy models like GHZ~\cite{ayral2021quantum} and also on problems of real-life interest such as finding approximate solutions to combinatorial optimization problems \cite{saleem2021quantum} or faster simulation of quantum circuits \cite{smith2023clifford}. In this study, we demonstrate the relevance of circuit cutting to exploit the use of smaller quantum machines, with the VQE as our use case. We have used the RealAmplitudes ansatz with a single repetition of the reverse-linear entanglement (a template for 3-qubits is shown in Fig.~\ref{fig:realamplitude}) which can be partitioned into two fragments using a single cut irrespective of the number of qubits. The fragments are executed parallely on two IBMQ devices, and the probability distribution of the uncut circuit is obtained using the Qiskit Circuit Knitting Toolbox \cite{circuit_knitting}.

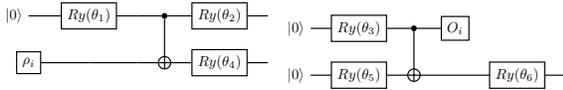
\begin{figure}[htb]
    \centering
    \begin{subfigure}{0.2\textwidth}
    \begin{adjustbox}{scale=0.55}
        \begin{quantikz}
            \lstick{$\ket{0}$}& \gate{Ry(\theta_1)} & \qw & \ctrl{1} & \gate{Ry(\theta_2)} & \qw \\
            \gate{\rho_i} & \qw & \qw & \targ{} & \gate{Ry(\theta_4)} & \qw \\
        \end{quantikz}
    \end{adjustbox}
    \end{subfigure}
    \begin{subfigure}{0.2\textwidth}
    \begin{adjustbox}{scale=0.55}
        \begin{quantikz}
            \lstick{$\ket{0}$}& \gate{Ry(\theta_3)} & \ctrl{1} & \gate{O_i} \\
            \lstick{$\ket{0}$}& \gate{Ry(\theta_5)} & \targ{} & \qw & \gate{Ry(\theta_6)} & \qw
        \end{quantikz}
    \end{adjustbox}
    \end{subfigure}
    \caption{Example of cutting the circuit of Fig.~\ref{fig:realamplitude}}
    \label{fig:cut}
\end{figure}

Fig.~\ref{fig:cut} shows the two fragments obtained by cutting the circuit of Fig.~\ref{fig:realamplitude} on the 2nd qubit between the two CNOT gates. $O_i$ and $\rho_i$ represent tomographically complete measurement and preparation bases respectively. Note that the total number of qubits from the two fragments is one more than that in the original circuit since one qubit, which is cut, is present in both the fragments \cite{peng2020simulating}.

In our experiment we cut a 6 qubit RealAmplitudes ansatz in two partitions consisting of 4 and 3 qubits, and report the expectation value for both the uncut circuit execution, and the two cut circuits execution with classical postprocessing. To the best of our knowledge, this is the first application of circuit cutting for estimating the ground state of a molecular Hamiltonian.

\subsection{Slicing the workload job batch} 
\label{sec:workfload-splitting:slice}
As discussed earlier, circuit cutting is a feasible option only when the circuit can be effectively partitioned into multiple subcircuits using a small number of cuts only. A QSVM circuit contains two-qubit interactions between all qubit pairs. Thus, basically every qubit has to be cut to create an effective partition. Circuit cutting, being infeasible in such a scenario, we shift our focus to data parallelization. This approach focuses on slicing the workload such that it can be executed with small resource foot-print or on low-capacity quantum hardware. Two possibilities include: (1) Splitting the workload into subsets which can be independently and parallely executed (2) Extracting a subset of the workload such that the objective of the application is achieved albeit with a smaller resource foot-print. 

We evaluate both these approaches in the case of QSVM kernel computation. For the first method, we split the circuit batch representing the data points of training and test sets such that they could be executed independently on multiple quantum hardware. The results for each of these batches are then combined to produce the kernel for the larger dataset. 

For the second method, we perform feature selection so as to find the most significant subset of features. Fewer features require fewer qubits and hence reduces the resource footprint of the workload.

These techniques could be potentially useful in applications such as QSVM where circuit cutting will lead to an unaffordable post-processing time due the complete graph connectivity between the qubits of the circuit.

\section{Evaluation Methodology}
\label{sec:evaluation}
\subsection{Workload Characteristics}

\subsubsection{VQE}
For our experimentation of a VQE workload, we consider the QHSM of Eq.~(\ref{eq:qhsm_zz}) with $n = 6$ (termed as $H_6$ henceforth). Using classical numerical methods we compute the minimum eigenvalue of $H_{6}$ to be $-5.0$ with a multiplicity of $2$. The two degenerate eigenstates are $\ket{\psi_{gs1}} = \ket{010101}$ and $\ket{\psi_{gs2}} = \ket{101010}$, which can be prepared via a depth-1 circuit starting from the initial state $\ket{000000}$. The objective of the VQE is to converge to a state such that the expectation value of the Hamiltonian $H_{6}$ is minimum. The Variational Principle guarantees that such a state is indeed the ground state for that Hamiltonian.

Since all the Pauli observable terms in $H_{6}$ commute with each other, and also with the observable $(\sigma^{z})^{\otimes 6}$, it is sufficient to measure all the qubits of the final state in the computational ($Z$) basis and use the resulting probability distribution to evaluate the expectation value for each of the terms in $H_{6}$. The \textit{Sampler Primitive} in Qiskit Runtime directly provides the probability distribution by measuring the quantum state in the $(\sigma^{z})^{\otimes 6}$ basis. 


When exploiting circuit cutting, the VQE ansatz is cut into 2 fragments (refer to Fig.~\ref{fig:realamplitude} and ~\ref{fig:cut}), each of which is evaluated independently and parallely with the other for a given set of parameters. The final expectation value is calculated via classical postprocessing, which is provided to the classical optimizer to search for better parameters. In other words, circuit cutting keeps the classical optimization method unchanged. It only affects the quantum portion where each fragment is executed on a smaller hardware - thus relaxing the number of qubits required. Moreover, since each fragment has lower qubit count and gate count, they are expected to be less susceptible to noise than the uncut circuit \cite{ayral2021quantum}.

\subsubsection{QSVM}
In QSVM, we construct a $n \times n$ training kernel matrix where we have $n$ training data points. Each entry of this kernel is obtained by running a circuit on the quantum hardware. This results in an execution of $n^2$ circuits for $n$ training data. In the IBM Quantum hardware these circuits are run in batches as jobs wherein each job can have a maximum of 300 circuits for larger devices ($>7$ qubits) and a maximum of 100 circuits per job in smaller devices. However while running these programs using Qiskit Runtime, we have been able to run more circuits per job on larger devices and we have also been able to parallelize job runs and then construct the kernel once all training jobs are completed. 

In this paper we used QSVM for classification of heart failure from an open-source dataset\footnote{https://archive.ics.uci.edu/ml/datasets/Heart+failure+clinical+records}, which contains clinical records of 299 patients with 13 clinical features for each patient \cite{HFdatapaper}. Out of the 299 patient data, 203 have survival and 96 have death outcomes labeled. As in any other machine learning task, we split the data to train and test partitions. We started with 80:20 train:test split but in order to train QSVM with 240 datapoints and test QSVM with the remaining 59 datapoints, we would need to run 28680 training circuits and 14160 test circuits. Since the larger machines have higher queue times, running such large number of circuits can take several weeks or error out while waiting on queue when run using the circuit API. So we randomly sampled 100 datapoints while maintaining the distribution of labels and used a 70:30 split for train and test. In this setup we needed to run 2415 training circuits and 2100 test circuits. When we run using circuit api we are restricted to 100/300 circuits in both the sequential and parallel runs. While in Runtime api, we have been able to run 1000 circuits per job for the 27 qubit devices (ibm\_montreal) and 600 circuits per job for 127 qubit device (ibm\_washington), thus reducing the number of jobs and improving overall compute time.

\begin{figure*}[htb]
  \centering
    \begin{subfigure}{0.4\textwidth}
       \includegraphics[width=\linewidth]{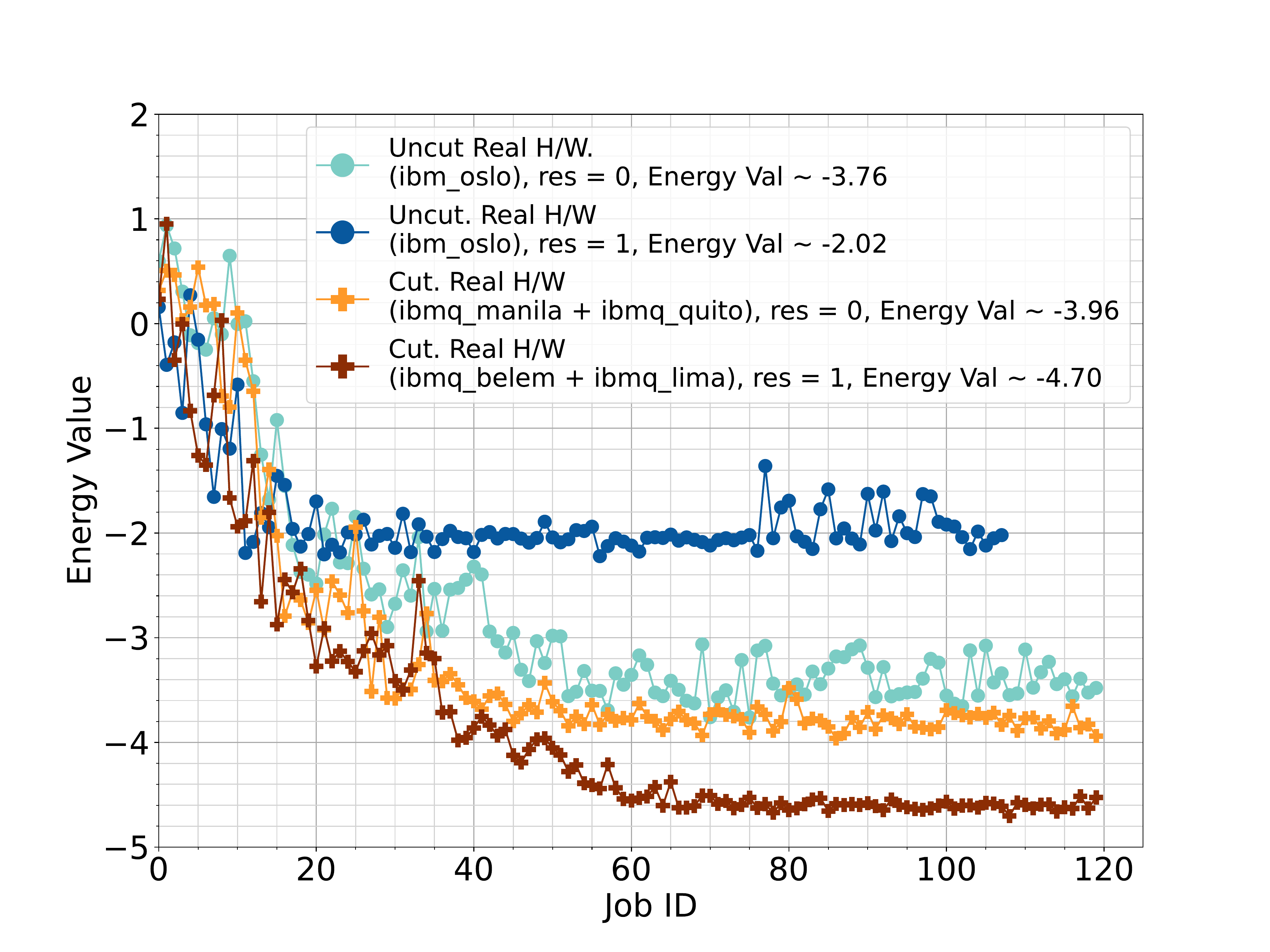}
         \caption{number of shots = 2000}
         \label{fig:cseq1000c6q}
    \end{subfigure}
    \begin{subfigure}{0.4\textwidth}
        \includegraphics[width=\linewidth]{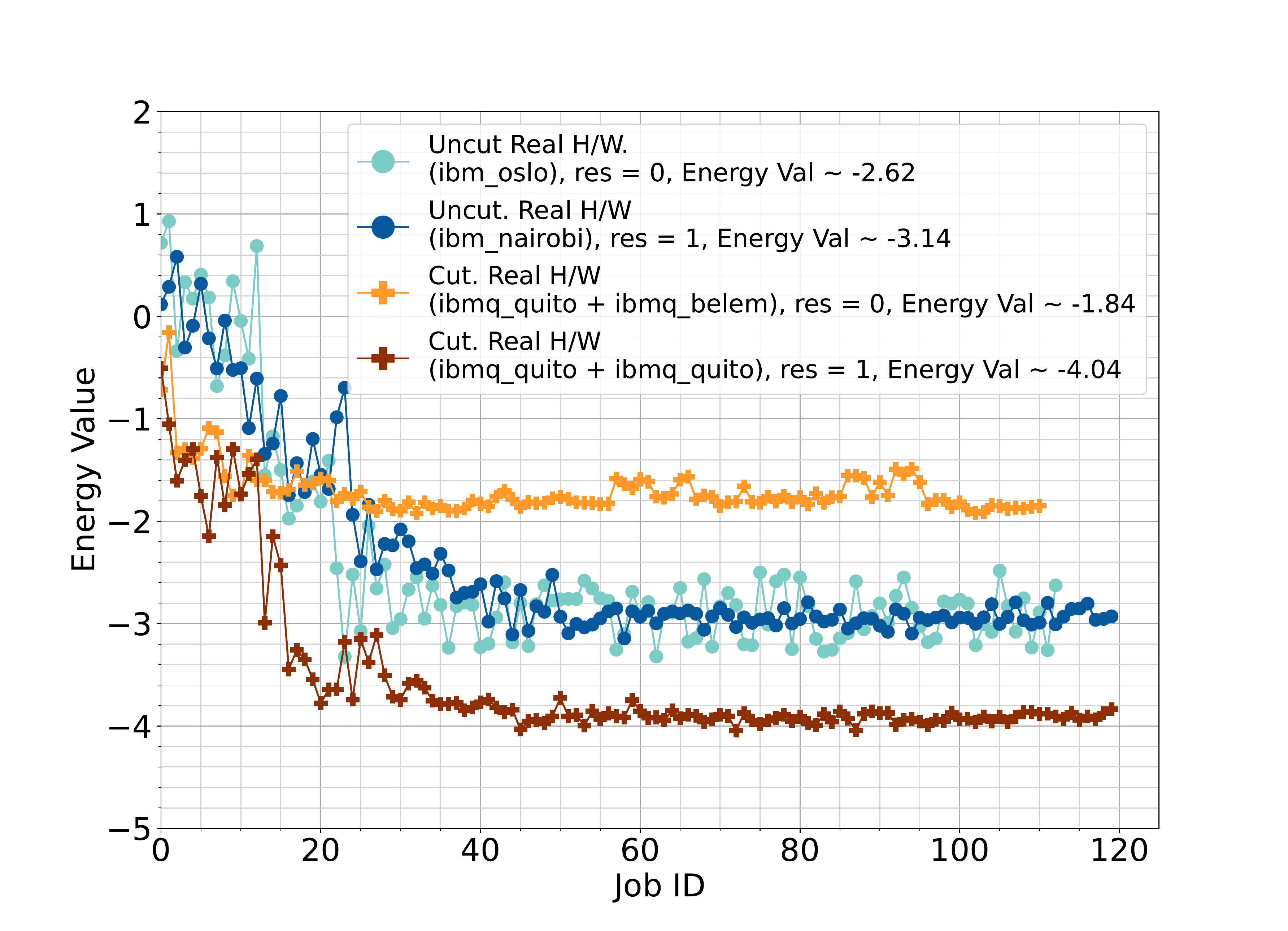}
        \caption{number of shots = 4000}
        \label{fig:cseq1000c6q}
    \end{subfigure}
    \begin{subfigure}{0.4\textwidth}
        \includegraphics[width=\linewidth]{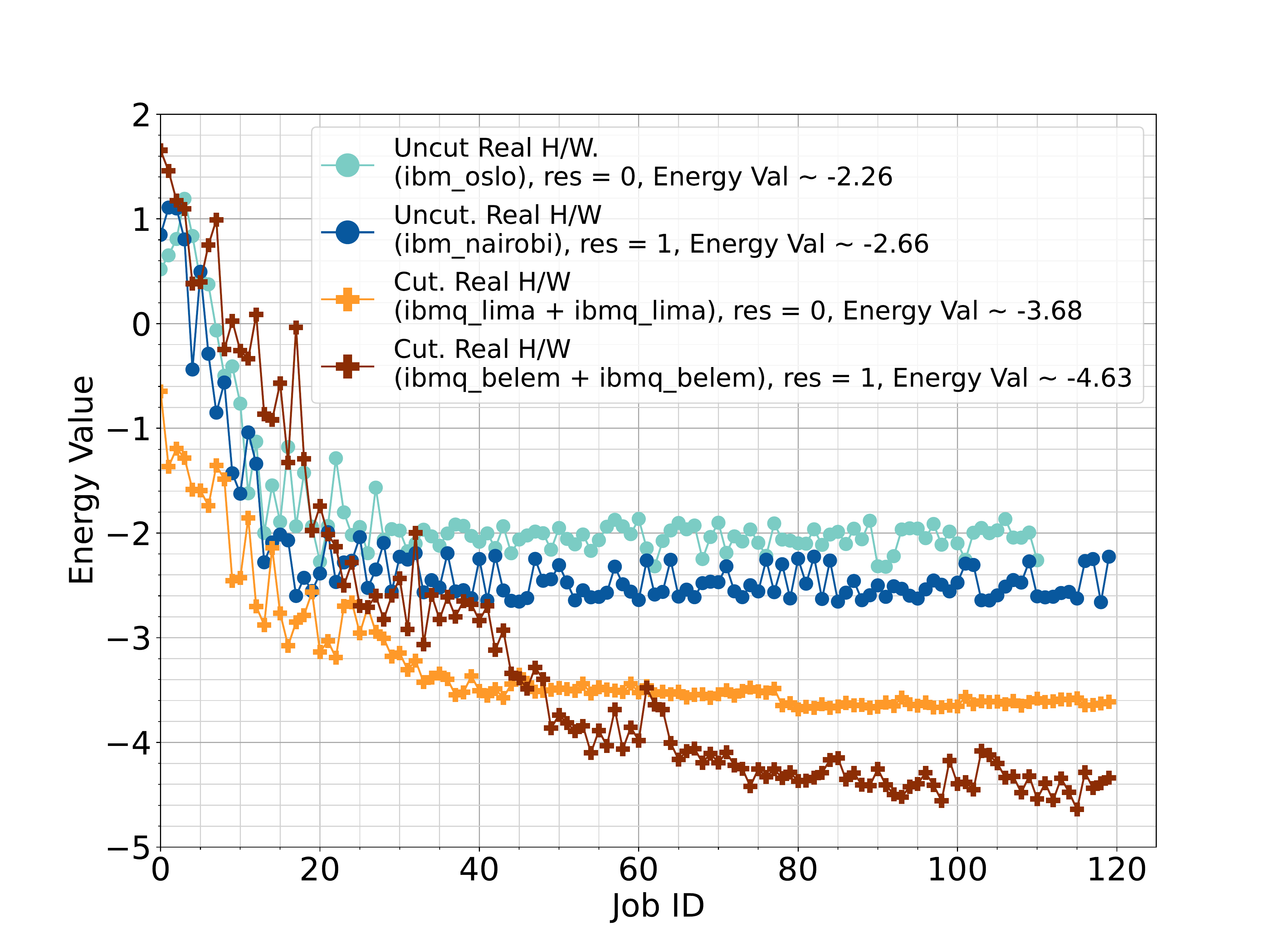}
        \caption{number of shots = 6000}
        \label{fig:cseq1000c6q}
    \end{subfigure}
    \begin{subfigure}{0.4\textwidth}
       \includegraphics[width=\linewidth]{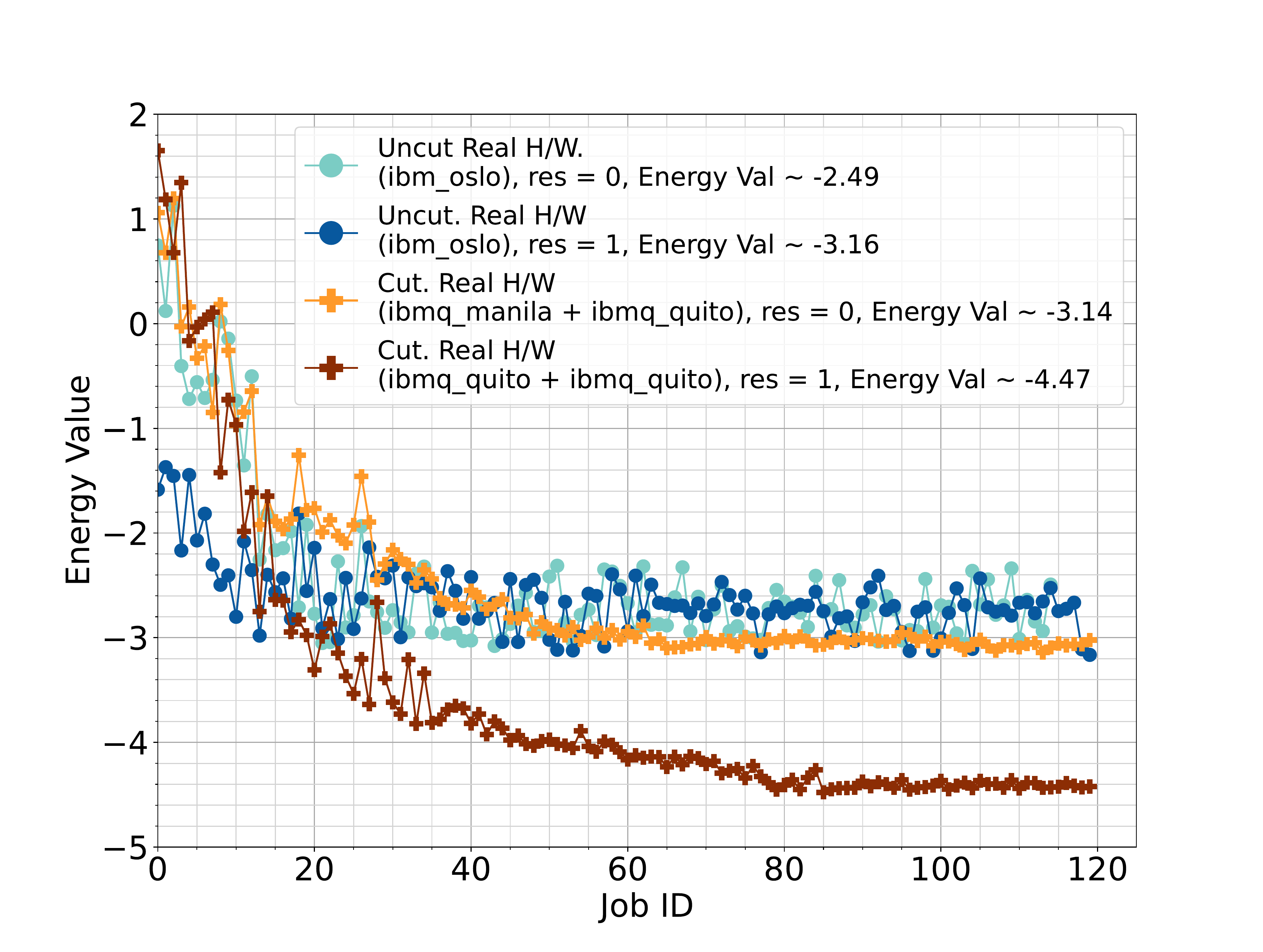}
        \caption{number of shots = 8000}
        \label{fig:cseq1000c6q}
    \end{subfigure}
    \caption{VQE performance using circuit-cutting on quantum hardware, with and without noise for different number of shots with resilience\_level 0 and 1}
    \label{fig:VQEcut_2k_4k_6K_8k_shots}
\end{figure*}

\subsection{Metrics} 
\label{sec:evaluation:metrics}
A baseline is required for each use-case for the sake of comparison. The uncut version of VQE forms the baseline for comparison with the cut versions. For QSVM, we use the sequential versions of the QSVM classifier as baseline to understand the impact workload slicing. Following are the metrics used for evaluation:
\begin{itemize}
    \item \textbf{Quality of result: } The quality of result for VQE is the energy level to which the algorithm converges. In case of QSVM, we report the accuracy and average Macro-F1 score of the classifier.
    \item \textbf{Resource foot-print: } The \emph{Qubit-Time product} captures the aggregate resource foot-print of the workload. A computation for $t$ time duration on $Q$ qubits has a qubit-time product of $Q \times t$.
    \item \textbf{Classical overhead: } This captures the total time spent in pre- and post processing the circuits. In the case of VQE, this will be the circuit cut and merge operations. In the case of QSVM, this will include the batching of circuits, and kernel merge operations.
    \item \textbf{Execution time: } This metric captures the maximum time taken for the batch of jobs to execute on the quantum machine. 
    \item \textbf{Pre-execution overhead: } Once a job is submitted to the quantum hardware backend, the circuits of the job are validated and queued until the backend hardware is available for execution. We capture this duration as the pre-execution overhead. Queuing delays might not even exist in all resource utilization models. For instance, reservation-based models will ensure little or no queues in an alloted user timeslot. Further, as the quantum hardware resources become more ubiquitous, queuing delays will play less of a role. Hence, we observe this metric mainly to understand if workload parameters affect queuing delay in current scenario.
    
\end{itemize}

\subsection{Instrumentation Methodology}
\label{sec:results:instrumentation_methodology}
To capture the client-side (the end-user application submitting the jobs), we build qiskit-runtime, circuit-knitting-toolbox, qiskit-terra and qiskit-aer from source and include instrumentation such as logging the JobID, Job Parameters, pre and post-processing timestamps. 

The job and backend characteristics for the IBM quantum cloud are retrieved using the IBM Qiskit Runtime API. This API provides the timestamps for state changes once the job is submitted to the platform. These states are labeled \emph{created}, \emph{running} and \emph{finished} to represent job creation, execution and completion events. Furthermore, the Runtime API also provides the quantum resource usage time in the \emph{quantum\_seconds} field of the job metadata.

The error mitigation technique to be used is set using the \emph{resilience\_level} field for the \emph{Sampler} and \emph{Estimator} primitive class of the Runtime API. The number of shots, seed, circuits to be used, observables and parameter vectors are also configured using the Runtime API.

In case of the circuit-cutting experiments, we also have modified the \emph{circuit-knitting toolbox} to efficiently manage the Runtime service session by persisting the sessions for the chosen backends and re-use the user-created session object. We achieve this by passing a custom session dictionary mapping the backend name to the session object into the \texttt{evaluate\_subcircuits()} function of the circuit knitting toolbox. The benefit of persisting the session (see Fig~\ref{fig:pre_exec_all_shots} related discussion) is that we get a fixed reservation for a specific time when the first job of the submitted program starts running due to which the subsequent jobs face no queueing delays.

\section{Results}

\subsection{VQE} In Fig.~\ref{fig:VQEcut_2k_4k_6K_8k_shots}, each sub-figure shows the VQE convergence behaviour curves of the expectation value of $H_{6}$ both \textit{with} and \textit{without} using circuit-cutting for different number of \textit{shots} for both resilience\_level 0 and 1 on real quantum hardware. 

We selected the \textit{least-busy} 5-qubit or 7-qubit machines having public access for the cut and uncut circuits respectively. In the circuit-cutting case, the fragments are executed parallelly, by having two active \texttt{Sessions}, on the two least-busy 5 qubit devices. The hardware(s) used to obtain the results are mentioned in the legend of the particular curve in Fig.~\ref{fig:VQEcut_2k_4k_6K_8k_shots} along with the energy value at which the VQE algorithm converges.

\begin{figure*}[htb]
  \centering
    \begin{subfigure}{0.4\textwidth}
         \includegraphics[width=\linewidth]{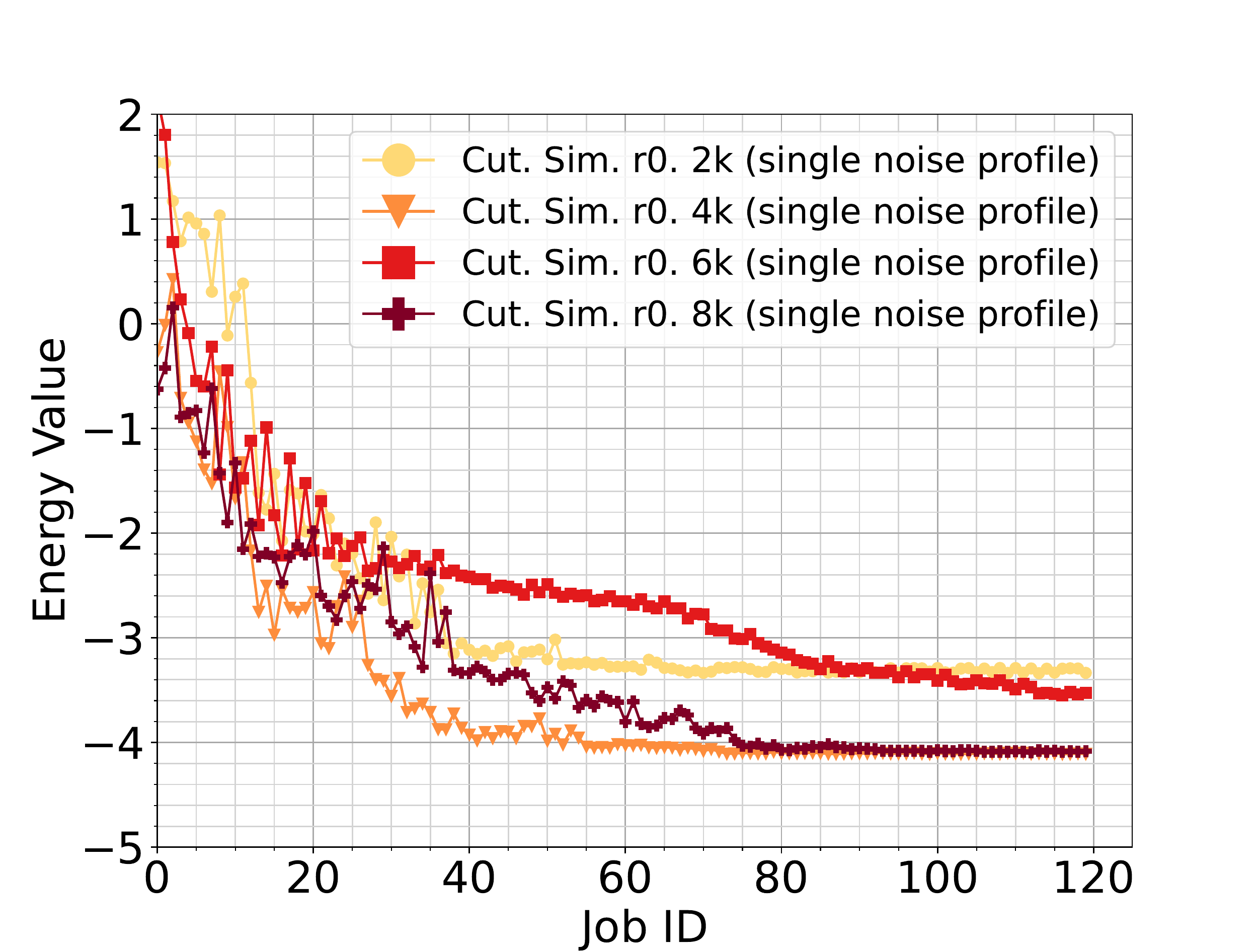}
         \caption{resilience\_level = 0}
         \label{fig:VQEcut_noise_r0_2k_4k_6K_8k_shots}
    \end{subfigure}
    \begin{subfigure}{0.4\textwidth}
         \includegraphics[width=\linewidth]{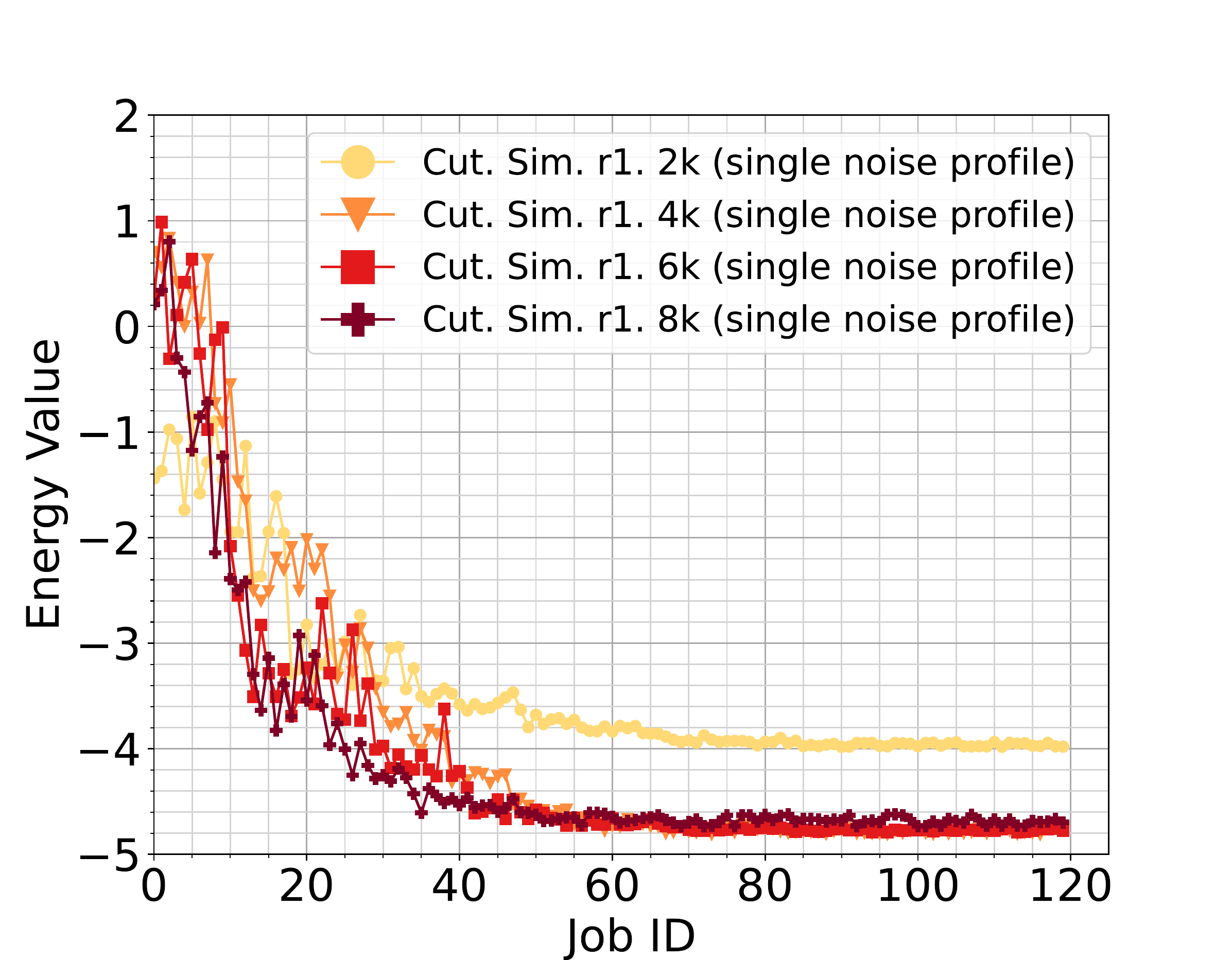}
         \caption{resilience\_level = 1}
         \label{fig:VQEcut_noise_r1_2k_4k_6K_8k_shots}
    \end{subfigure}
    \caption{Comparing VQE with circut-cutting for different number of shots: (a) with and (b) without error mitigation simulated using two different snapshots of the noise profiles of \textit{ibmq\_belem} and \textit{ibmq\_lima}}
    \label{fig:VQEcut_noise_2k_4k_6K_8k_shots}
\end{figure*}

\begin{figure}[h!]
  \centering
         \includegraphics[width=0.9\linewidth]{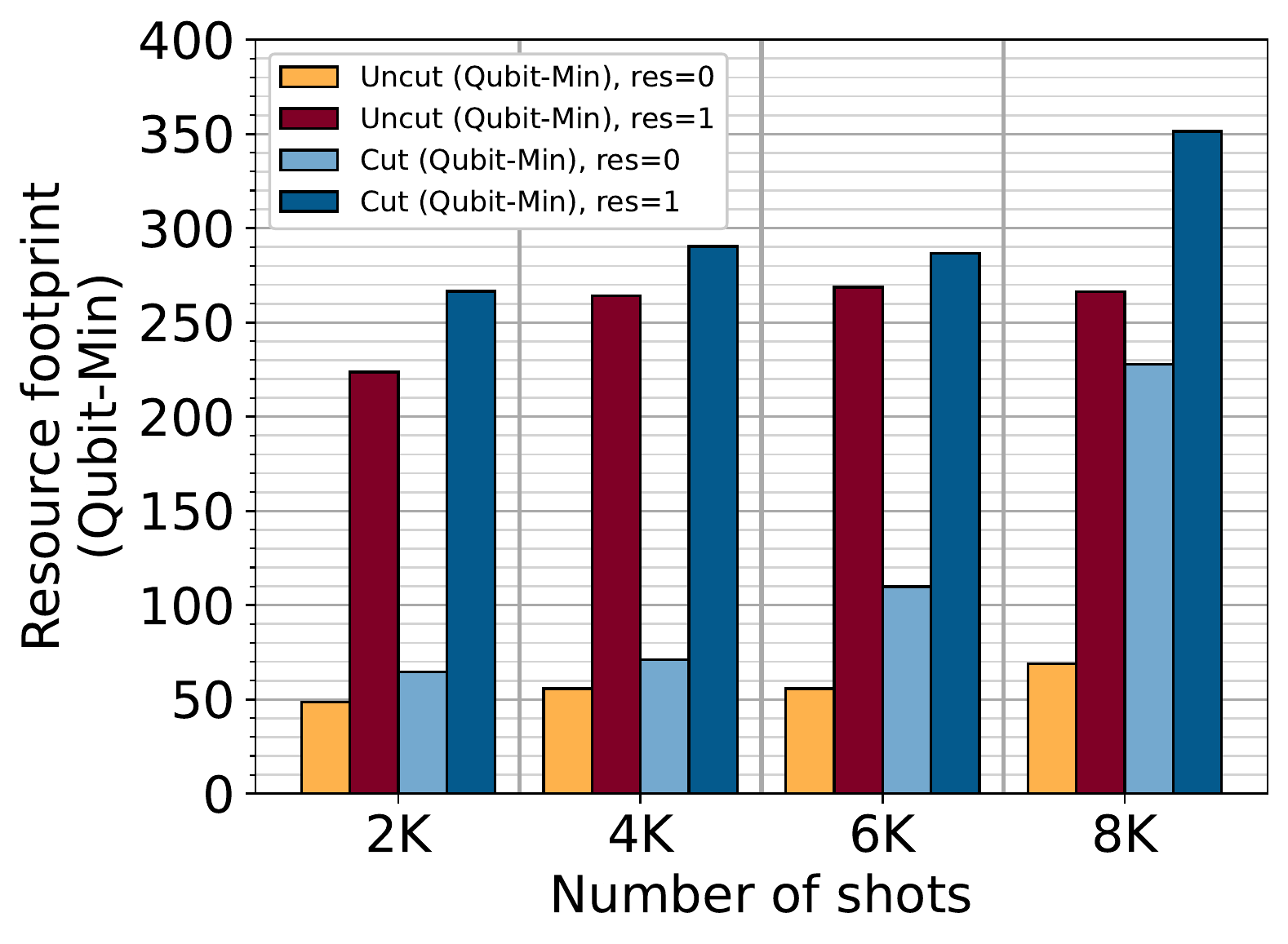}
         \label{fig:VQE_exectime_qrusage_r0_cut_vs_uncut}
    \caption{Variation of Resource footprint for different shot configurations with and without circuit-cutting}
    \label{fig:VQE_resource_footprint}
\end{figure}

\textbf{Analysis of Quality of Energy Convergence:}
Circuit cutting involves measurment and preparation in different basis, and is, therefore, highly susceptible to state preparation and measurement (SPAM) error \cite{majumdar2022error}. Therefore, we find the expectation values of the cut and uncut versions for both resilience\_level 0 (no mitigation) and 1 (measurement error mitigation). In general, we note that the expectation value obtained using circuit cutting is even closer to the optimal value than the uncut one for all range of shots. Naturally measurement error mitigation provides a result closer to the optimal value; but even there circuit cutting is shown to outperform the corresponding uncut version. We note from the result that the number of steps required for convergence is higher for the cut versions than the uncut one. Nevertheless, this is outweighed by the $\sim 39\%$ improvement in the obtained result provided by circuit cutting. Noise in each fragments is lower than that of the uncut circuit due to lower qubit count, lower depth and lower gate count. Our observation suggests that the classical optimizer, in case of circuit cutting, can find parameters \emph{closer} to the optimal ones than for the uncut scenario. This, sometimes, leads to a slower convergence for circuit cutting since here the optimizer can find better parameters than the uncut one.

From Fig.~\ref{fig:VQEcut_2k_4k_6K_8k_shots} we observe some discrepancy in the reported energy values. For example, the energy value is expected to be a non-decreasing function of the number of shots. However, for example, 4000 shots seem to provide a result worse than 2000 shots. In some cases, hardware result with resilience\_level 1 fails to outperform that with resilience\_level 0 for the uncut version. Recall that we selected two \emph{least-busy} devices to execute the circuits. As a result, the hardwares on which the fragments are executed sometimes change between the different shot instances making our results subject to hardware heterogeneity arising from the public access model. Moreover, the queuing time prior to execution varies for different hardware and instances. Furthermore, transpilation in qiskit involves stochastic steps. This can end up choosing noisier qubits in some transpilation step for some execution cycle. Also, the noise profile varying with time is an unavoidable consequence of the changes in the microenviroment of the quantum hardware. 

As an empirical investigation whether the above suggested reasons are indeed responsible for the observed deviations from the expected trend, we look at the VQE behaviours using circuit-cutting for noisy simulation using the noise models of \textit{ibmq\_belem} and \textit{ibmq\_lima} for both resilience\_level 0 and 1 in Fig.~\ref{fig:VQEcut_noise_2k_4k_6K_8k_shots}. We find that the expected trend of better convergence with increasing shots is observed in the simulation results {using a fixed noise model.

Circuit cutting is primarily subjected to state preparation and measurement (SPAM) error due to the multiple measurement and preparation basis required \cite{majumdar2022error}. Moreover, the circuits are also small in our use-cases and we obtain sufficiently good results with TREX mitigation only. Using ZNE will incur higher resource-footprint and therefore we do not use it for this study.

\begin{figure}[htb]
  \centering
    \includegraphics[width=0.95\columnwidth]{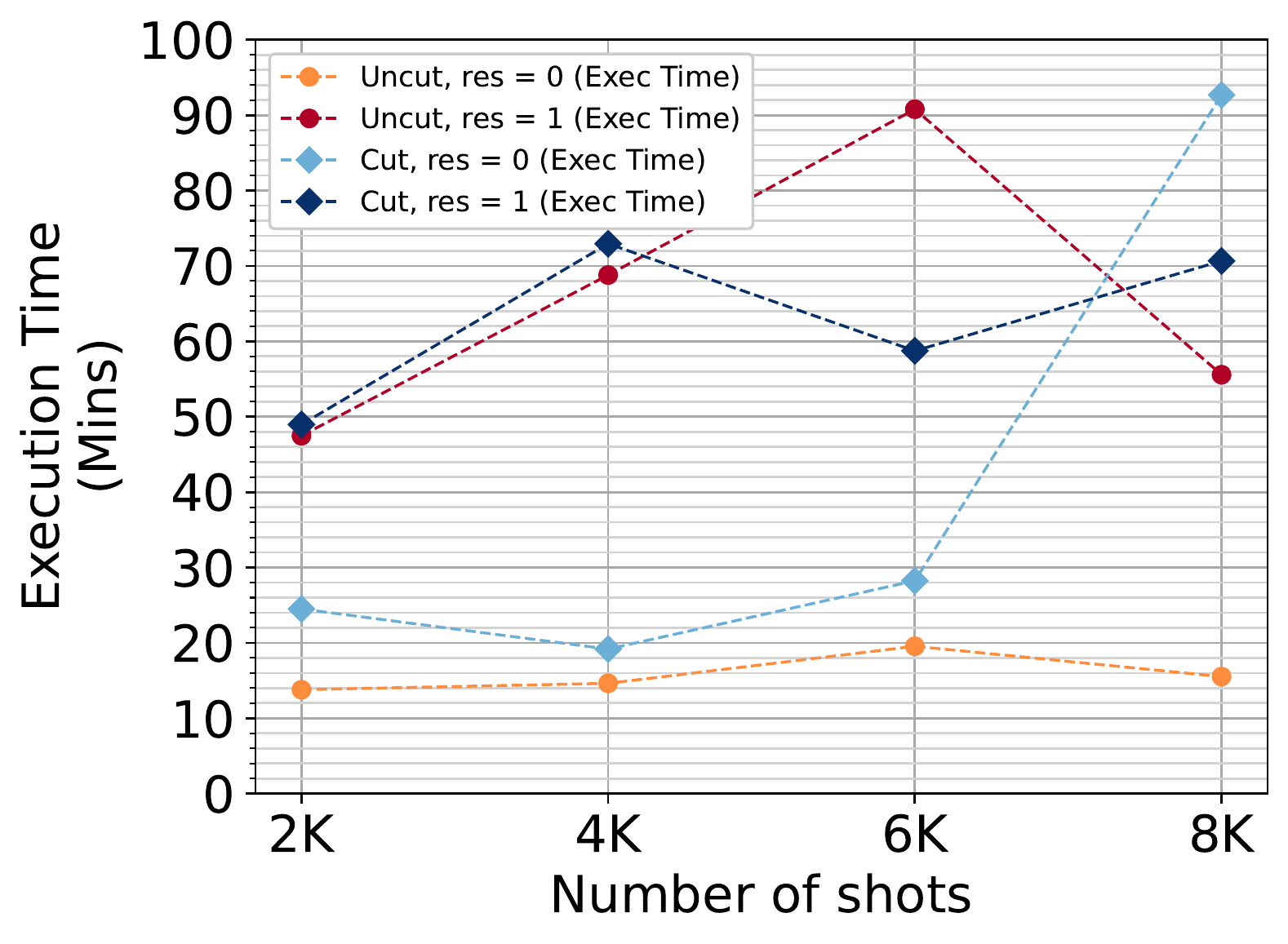}
    \caption{Execution Time for different VQE instances}
    \label{fig:execution_time_cut_uncut}
\end{figure}

\begin{figure}[htb]
  \centering
    \includegraphics[width=0.95\columnwidth]{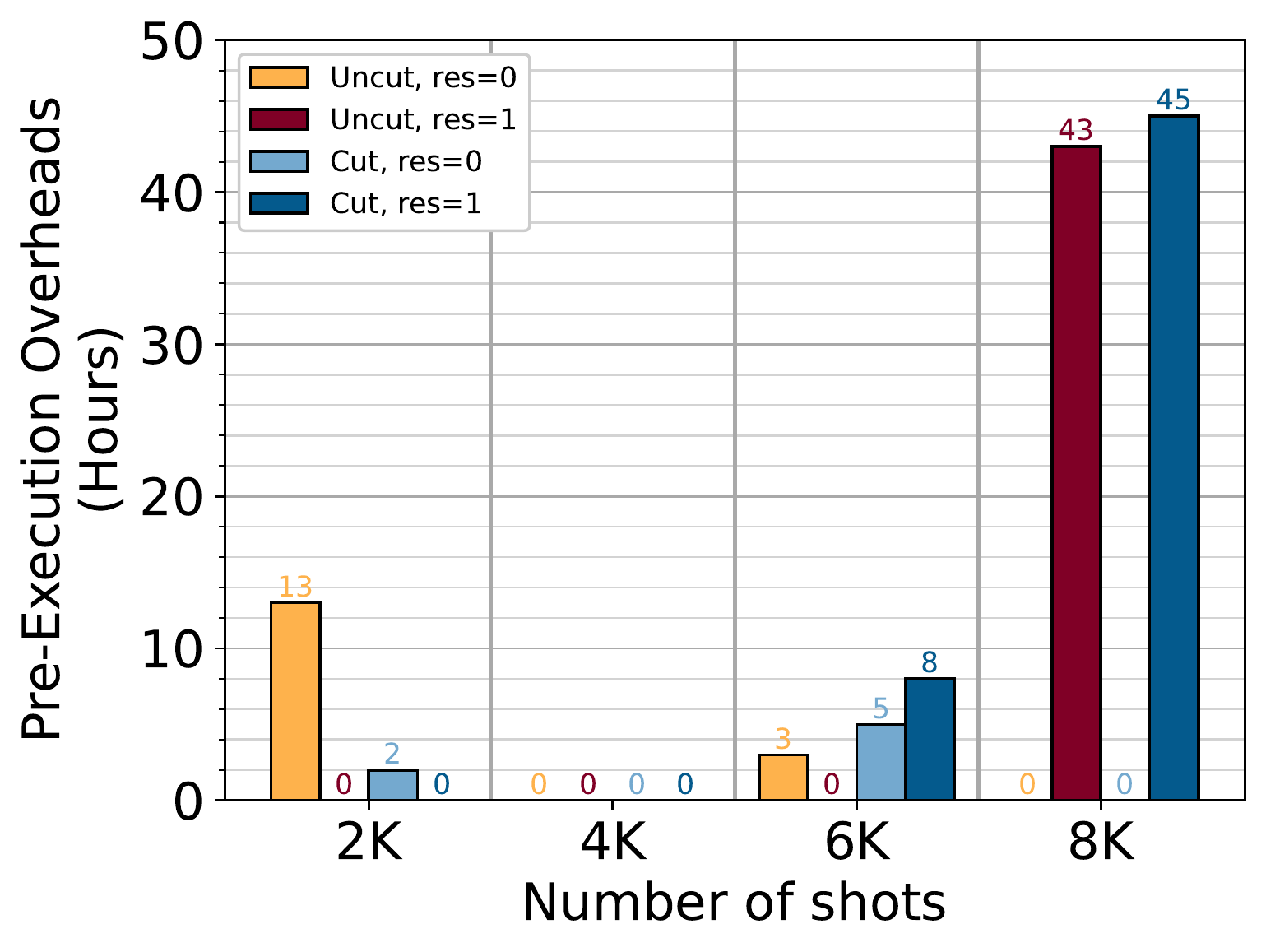}
    \caption{Pre Execution Overheads for VQE with circuit splitting on ibmq\_belem and ibmq\_quito at 4k shots}
    \label{fig:pre_exec_all_shots}
\end{figure}

\textbf{Resource foot-print:} The bar plot in Fig.~\ref{fig:VQE_resource_footprint} captures the resource usage by the VQE instances using the qubit-minute metric. The details of the \textit{Sampler Primitive} jobs can provide the time spent by both the quantum resource and the classical resource (Sec.~\ref{sec:results:instrumentation_methodology}). As mentioned in Sec.~\ref{sec:workfload-splitting:ckt-cut}, the circuit-knitting toolbox cuts the 6-qubit circuit into a 4-qubit circuit and a 3-qubit circuit. Therefore, in every individual VQE iteration using circuit cutting, one job executes a 4-qubit circuit while the other executes a 3-qubit circuit. The the total \textit{Resource foot-print} for a iteration will then be the sum of the resource foot-print for the individual jobs. We consider the \textit{Resource foot-print} of the the case without circuit-cutting as the baseline case. From Fig.~\ref{fig:VQE_resource_footprint} we observe that the resource-footprint of circuit cutting is always greater than that of the uncut circuit for both the resilience levels. Note that the definition of \textit{resource foot-print} (Sec.~\ref{sec:evaluation:metrics}) captures the aggregate time resources (upto a factor of the number of qubits used) consumed by a VQE instance, regardless of whether the resources were used parallely or not. In aggregate circuit cutting requires one qubit more than the uncut one (refer to Fig.~\ref{fig:cut}). Also, as discussed earlier, it has a slower convergence than the uncut one but it converges to a more accurate energy value. Both of these contribute to the higher resource-footprint for circuit cutting.

\textbf{Execution time:} Fig.~\ref{fig:execution_time_cut_uncut} reports the execution time of the uncut and cut circuits. For cut circuits we report the maximum of the execution time of the two fragments since they are executed parallely in two different hardware. The execution time of the cut circuits are, in general, higher than that of the uncut one, except for an anomaly for 6000 shots with resilience\_level 1. This higher execution time is due to (i) the slower convergence of the cut circuits, and (ii) the lower CLOPS value of the 5-qubit devices (see Table~\ref{tbl:machineprofile}). CLOPS is a metric which captures the execution speed of a quantum hardware \cite{wack2021scale}. Since most of the 5-qubit devices are slower than the 7-qubit ones, the cut circuits naturally require higher execution time. Nevertheless, since the cut circuits are smaller than the uncut ones, the execution times for the two cases do not differ significantly. Cut with non resilience gives comparable or better results than the uncut versions for 2000 and 6000 shots while also taking much less (approximately half) time than scenarios with error mitigation.

Measurement error mitigation involves running multiple twirled circuits \cite{van2022}, and postprocessing the observed value to mitigate the effect of SPAM error. Since the number of circuits to be executed is higher, we always observe that the execution time with resilience\_level 1 is higher than that for resilience\_level 0.

\textbf{Pre-execution overhead:} Qiskit Runtime service~\cite{QiskitRuntime} sessions only suffer queuing delay per session. Once the machine is allocated to a given session, the jobs submitted to that session get executed with minimal delay. Our measurements indicate that the VQE jobs get scheduled for execution within 35 seconds on an average, once the session acquires the machine. From Fig.~\ref{fig:pre_exec_all_shots}, we also notice that the the pre-execution overhead which includes the queuing delay for the session is not influenced by the number of shots selected for the experiments. For instance, the sessions for 4000 shots suffer insignificant ovehead compared to 2000 shots. It appears this overhead is more dependent on the availability of machines at the time of job submission than the shots parameter.

\textbf{Classical overheads:} Circuit cutting involves classical pre and post-processing respectively to cut the original circuit into sub-circuits and calculate the final expectation value from the fragment outcomes \cite{peng2020simulating, tang2021cutqc, majumdar2022error}. This incurs some increased classical time requirement which is absent in the uncut version. However, we find that the classical pre and post processing times are minimal and are significantly less than the execution time. On an average over the number of \textit{shots}, the classical pre-processing overhead (transpilation, parameter binding, etc.) \textit{without circuit-cutting} is $\approx 3.02s$ and $\approx 2.78$ for resilience\_levels 0 and 1 respectively. Whereas, that \textit{with circuit-cutting} is $\approx 10.45s$ and $\approx 11.11s$. Further, the classical post-processing overhead \textit{with circuit-cutting} are $\approx 1550s$ and $\approx 1400s$ for resilience\_levels 0 and 1 respectively.

\subsection{QSVM} We evaluate ten workload processing approaches to compute the QSVM kernels. Each workload varies the number of features used, number of circuits per job, sequential/parallel execution and error mitigation technique used. All the workloads have a training and testing kernel computation, circuit batch splitting, model fitting and prediction components and kernel merging components.

First, we describe the workload types which do not use any error mitigation techniques:

\texttt{6Q-1000C-SEQ: }This type represents a QSVM classifier that leverages the Runtime API and the corresponding \emph{Estimator} primitive to compute the QSVM kernels. \emph{1000 circuits} are packed per job and \emph{6 features} are selected from the original dataset, thus 6 qubit circuits running their jobs sequentially on \emph{ibm\_oslo}.

\texttt{6Q-1000C-PAR: }This type parallelizes 6Q-1000C-SEQ by executing the testing and training kernel computation as separate simultaneously running processes. Within each phase, the jobs execute sequentially. This particular workload organization is meant to factor in the limits imposed by the quantum cloud platform on a number of parallel jobs submitted by a given user. This also attempts to explore the benefits of batching jobs within parallel executing processes.  ibmq\_nairobi and ibm\_oslo are used to run the jobs.

\texttt{6Q-500C-SEQ: }This is similar to 6F-1000C-SEQ but the number of circuits per job is reduced to 500, which in turn doubles the number of jobs. We specifically wanted to explore this setting for ablation studies, also the default max circuits per job is 300 so both the settings are improvements that we try to do to run QSVM on more data faster. This type was run on ibm\_oslo. 

\texttt{6Q-500C-PAR: }This type parallelizes 6Q-500C-SEQ using the same method as 6Q-1000C-PAR, and used the backends ibmq\_nairobi and ibm\_hanoi.

\texttt{2Q-500C-SEQ: }This type explores the possibility of feature selection by reducing the number of features (qubits) required to 2. This type is similar to 6F-500C-SEQ, but explores the impact of circuit width on performance. We use ibmq\_lima to execute this workload.

\texttt{2Q-500C-PAR: }This type parallelizes 2F-500C-SEQ along the same lines as 6Q-500C-PAR. ibm\_lima and ibmq\_manila are used to execute the workload.

We explore potential benefits of improving the accuracy even when using fewer qubits, and the corresponding trade-offs involved using the following workload types. We used ibmq\_lima, ibmq\_manila, ibm\_oslo, ibmq\_belem machines for below experiments:

\texttt{2Q-500C-SEQ-TREX: }This is the same workload type as 2Q-500C-SEQ, except that while initializing the Estimator class, we set the error mitigation parameter (\emph{resilience\_level = 1}) to use measurement error mitigation~\cite{van2022}. 

\texttt{2Q-500C-SEQ-ZNE: } This is the same as workload type 2Q-500C-SEQ, executed with ZNE error mitigation~\cite{temme2017} (\emph{resilience\_level = 2}). 

\texttt{2Q-500C-SEQ-TREX} and \texttt{2Q-500C-PAR-ZNE: } parallelize their sequential counter-parts 2Q-500C-SEQ-TREX and 2Q-500C-SEQ-ZNE respectively along the same lines as 6Q-500C-PAR.

\begin{figure}[!h]
     \centering
         \includegraphics[width=0.8\linewidth]{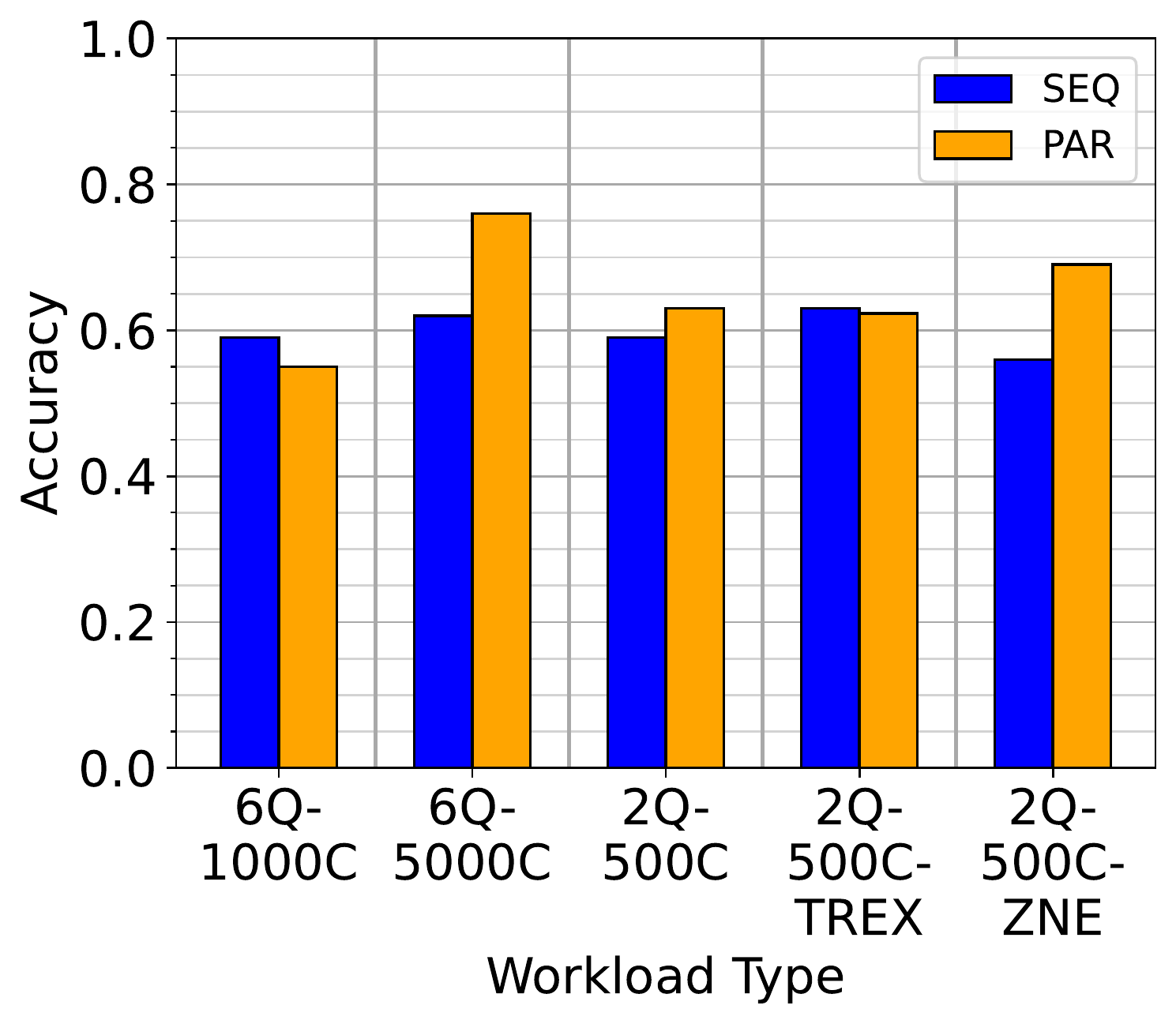}
         \caption{QSVM prediction accuracy for all the workload types}
         \label{fig:accuracy}
\end{figure}

\textbf{Accuracy. }Figure~\ref{fig:accuracy} illustrates the accuracy values for the various workload types comparing the sequential and parallel versions. The maximum accuracy reported among all the scenarios is for 6Q-500C-PAR. When the number of features used is reduced from 6 to 2, while keeping the number of circuits constant (500 circuits), the accuracy drops by 5\% for the sequential and 17\% for the parallel scenarios. While this is expected, the application of TREX error mitigation yields marginal benefit, but ZNE error mitigation improves the accuracy of the parallel version yielding 9\% improvement compared to the 2Q-500C-PAR scenario with no error mitigation. Compared to the baseline 6Q-500C-SEQ scenario, reducing the qubit requirement to 2, applying parallelization and ZNE error mitigation yields 11\% improvement. An important observation while realizing the accuracies achieved by all sequential runs with and without error mitigation is that the accuracies do not drastically improve with error mitigation, or the model was able to learn well inspite of the hardware noise. This is the why we didnt experiment with complex error mitigation strategies like PEC\cite{van2022probabilistic}. 

We conjecture that the improvements that we see for parallelized scenarios could be because the parallel versions leverage multiple machines which could have differing noise profiles. While a sequential version is committed to execute all its jobs on a single machine which if having a bad noise profile could affect all the jobs, the parallelized version could leverage the differences in noise profile across machines to execute some of jobs in a relatively better noise profile. Secondly, we also notice workload types using 500 circuits (compared to 1000 circuits) yielded better or very close accuracy values for parallelized versions compared their sequential counter-parts. We speculate that smaller batches of circuits could mitigate accumulated estimation errors for the estimator primitive. Further investigation might be required to substantiate this conjecture. 

\begin{figure}[!h]
     \centering
         \includegraphics[width=0.8\linewidth]{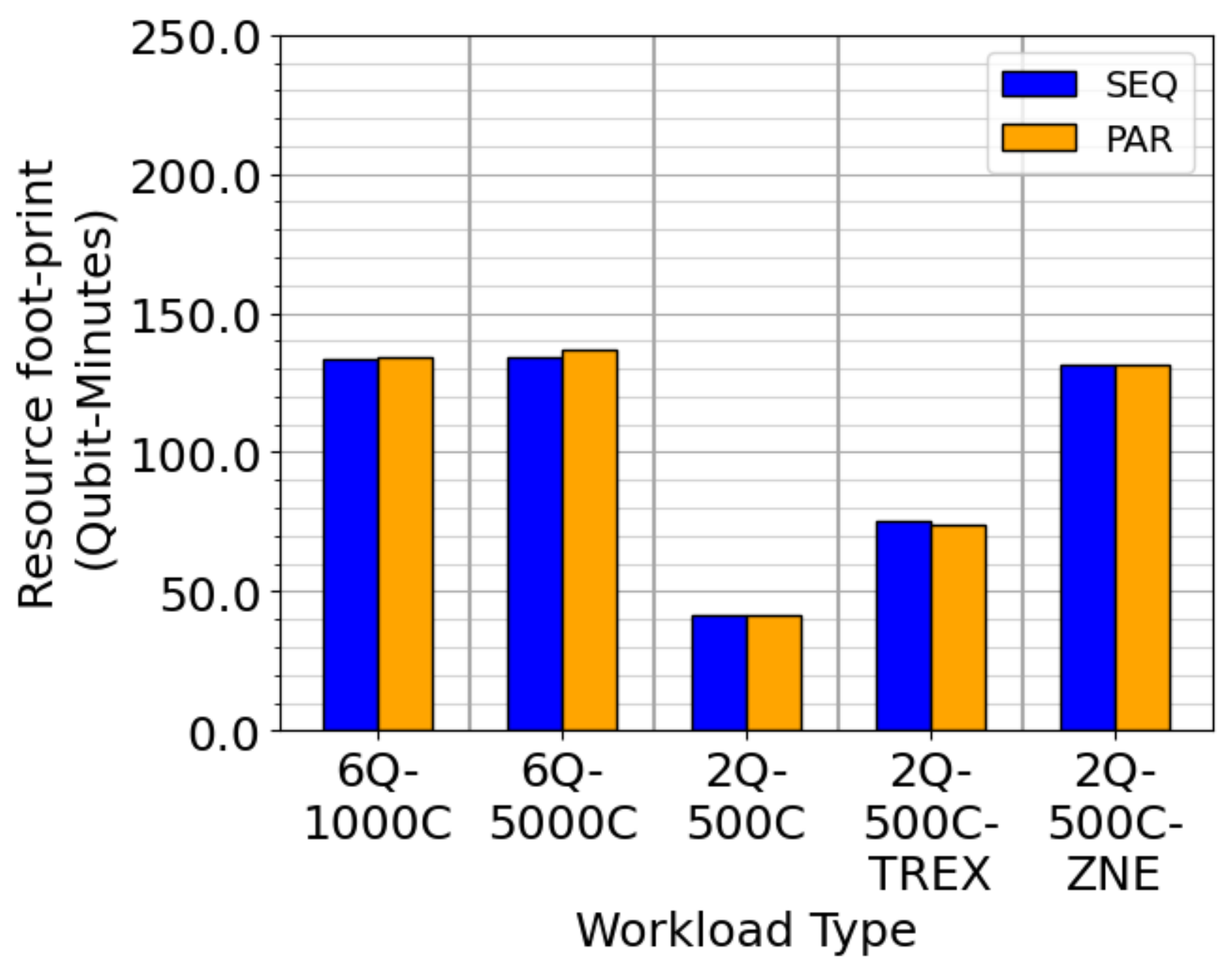}
         \caption{Resource foot-print of jobs for QSVM workload types}
         \label{fig:resfp}
\end{figure}

\textbf{Resource foot-print. }Figure~\ref{fig:resfp} captures the resource usage using the qubit-minutes metric. The first improvements in resource foot-print are derived from reducing the number of qubits used through selecting subset of features. When the baseline case of 6Q-500C-SEQ is compared with the 2Q-500C-SEQ, a 3x reduction in resource foot-print is observed. However, the accuracy also drops by 4\%. The parallel version 2Q-500C-PAR also provides a 3x reduction in resource foot-print, but also provides a comparably better (4\% improvement) accuracy to the baseline 6Q-500C-SEQ. The improvements in accuracy could be attributed to same reasons discussed in the previous sections. While, error mitigation improves the accuracy even further to nearly 7\% of the best case (6Q-500C-PAR), it also reduces the gains in the resource foot-print yielding 4\% improvement compared to 6Q-500C-PAR. This reduction is attributed to increased quantum circuit complexity to mitigate the errors resulting in higher execution times on the quantum hardware.

\begin{figure}[!h]
     \centering
         \includegraphics[width=0.8\linewidth]{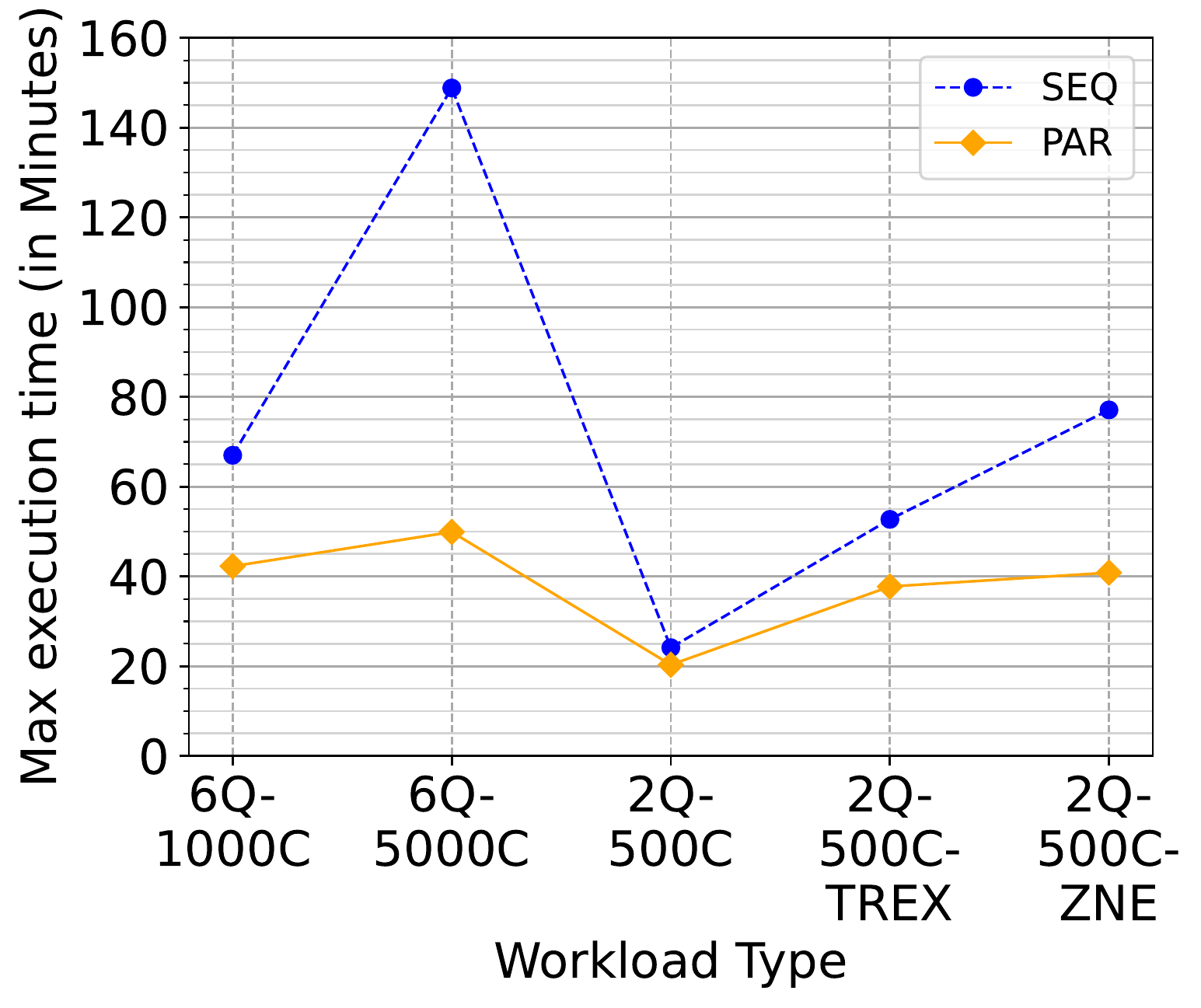}
         \caption{Execution times for the QSVM workload types}
         \label{fig:exectime}
\end{figure}

\textbf{Execution time. } Figure~\ref{fig:exectime} illustrates the execution time for the various QSVM workload types. Firstly, the gains in parallelization (obtained by comparing the sequential and parallel versions of each workload type) vary from 1.2x-3x. Speed-up due to simply using fewer qubits is 6x (obtained by comparing 2Q-500C-SEQ and 6Q-500C-SEQ). Adding parallelization to this qubit reduction yields a further 15\% improvement. Error mitigation techniques increase the execution time. However, coupled with parallelization and reduced number of qubits, 2Q-500C-PAR yields 11\% better accuracy than 6Q-500C-SEQ (which does not use any of the said techniques), while also computing the results 72\% faster. The other observation which is evident from Figure~\ref{fig:exectime} is that the execution time approximately doubles when the number of circuits per job is reduced from 1000 to 500. This could be due to the increased per job overhead that adds to the aggregate execution times for a given batch of jobs.

\textbf{Pre-execution overhead. } Figure~\ref{fig:preexectime} illustrates a cumalative distribution of the observed pre-execution overheads for jobs generated when workloads of type 6Q-100C, 6Q-500C and 2Q-500C were submitted for execution. We observe that 70\% of the jobs using fewer qubits (2Q-500C) tend to have pre-execution overhead of less than 100 minutes, and the remaining 30\% tend to have 500 minutes or less of overhead. Compared to this, 70\% of the jobs using 6 qubits (6Q-500C) and 500 circuits per job tend to 250 minutes of overhead or less. About 30\% of the jobs tend to have overhead between 100 minutes to 250 minutes, and another 30\% tends to have extreme overheads of 3000 minutes. Using more qubits and having more jobs (fewer circuits per job) tends to expose the workload to the variations of queuing behavior observed on current quantum cloud platforms with limited resources and large global user base. Reducing the number of jobs (more circuits per job) improves the situation relatively by reducing the overhead to 1000 minutes or less for 25\% of the jobs as can be seen in the case of 6Q-1000C.
\begin{figure}[!h]
     \centering
         \includegraphics[width=0.8\linewidth]{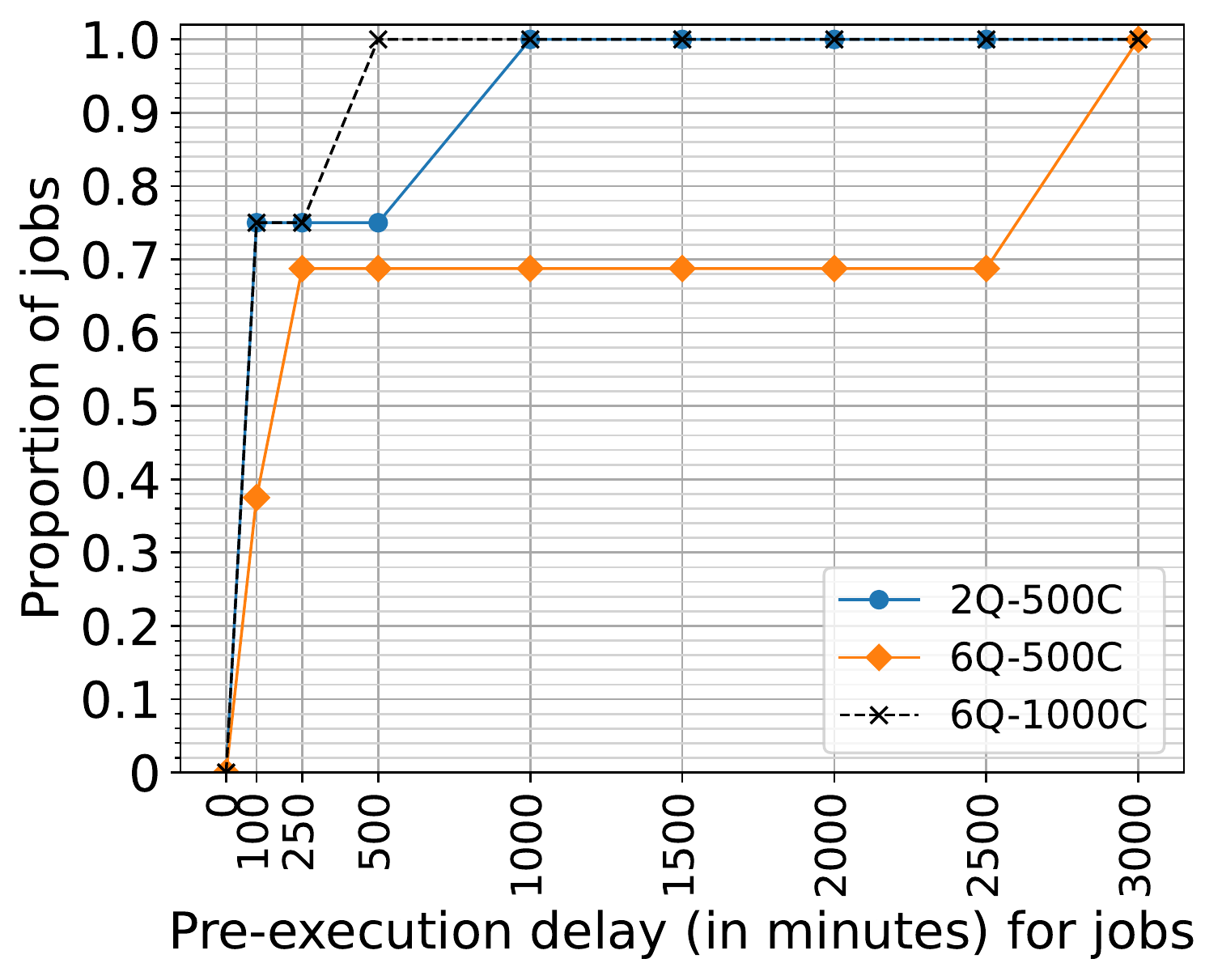}
         \caption{CDF of pre-execution overhead for all the jobs constituting the QSVM scenarios}
         \label{fig:preexectime}
\end{figure}

\textbf{Classical processing overheads. }The classical processing overhead captures the time taken to create the job batches, use the quantum job results to compute the training test kernels for the parallelized workload scenarios. This overhead 50 milliseconds to 107 milliseconds, with 2F-500C scenario having the lowest overhead, while 6F-500C have the highest. The number of features used and the number of circuits per job impact the number of iterations to compute the job batches and the kernels resulting in this difference in classical processing overhead.

\section{Discussion and Conclusions}
\label{sec:conclusion}
In this study, we profile two application use-cases representing real-world problems on real quantum hardware backends of diverse capacity. We explore the possibility of designing smaller, parallelizable hybrid quantum-classical jobs using workload slitting techniques. We also explore and analyze the trade-off between quality of the result of the application, the performance of the application and the reduction in resource foot-print. Our observations support the need for dynamic job orchestration platform to effectively and affordably leverage the diverse quantum hardware and address the challenges of limited availability, to effectively exploit the dynamic nature of the changes to hardware characteristics (e.g. noise profile, queues). While, Stein et al~\cite{eqc} have begun this process by proposing to exploit differences in noise profiles to achieve faster VQA convergence without actually adapting the workload, our study explores the workload and application-centric angles when even the workload could be adapted as follows:
\begin{enumerate}
    \item \textbf{Coupling workload adaptation and parallelization: } 
    Parallel execution of job batches could help to leverage additional but diverse capacity as it emerges. But, the workload itself might not be directly amenable to exploit this opportunity. Circuit cutting or feature sub-selection could help adapt the workload and also leverage parallelization. Without dynamic orchestration of this nature, related jobs running on different machines could become a bottleneck subject to changing machine state.
    \item \textbf{Failure Tolerance: } Jobs could fail because of various reasons (e.g. classical out-of-memory issue during heavy error mitigation and post-processing, network disruption between the client-server and the quantum hardware). Orchestration platforms should dynamically adapt to such points of failure and be designed to create redundancy within user budgetary constraints.
    \item \textbf{Judicious use of error mitigation: }Our profiling study has demonstrated the improvement in quality of result for both VQE and QSVM when error mitigation is used. But, results indicate that it also comes with increased utilization of quantum resources. Under budgetary constraints, the orchestration platform could selectively utilize error mitigation while being aware of resource costs.
\end{enumerate}

\bibliographystyle{IEEEtran}
\bibliography{main}

\begin{thebibliography}{10}
\providecommand{\url}[1]{#1}
\csname url@samestyle\endcsname
\providecommand{\newblock}{\relax}
\providecommand{\bibinfo}[2]{#2}
\providecommand{\BIBentrySTDinterwordspacing}{\spaceskip=0pt\relax}
\providecommand{\BIBentryALTinterwordstretchfactor}{4}
\providecommand{\BIBentryALTinterwordspacing}{\spaceskip=\fontdimen2\font plus
\BIBentryALTinterwordstretchfactor\fontdimen3\font minus
  \fontdimen4\font\relax}
\providecommand{\BIBforeignlanguage}[2]{{%
\expandafter\ifx\csname l@#1\endcsname\relax
\typeout{** WARNING: IEEEtran.bst: No hyphenation pattern has been}%
\typeout{** loaded for the language `#1'. Using the pattern for}%
\typeout{** the default language instead.}%
\else
\language=\csname l@#1\endcsname
\fi
#2}}
\providecommand{\BIBdecl}{\relax}
\BIBdecl

\bibitem{aws_quantum}
``{Amazon Braket, AWS},'' \url{https://aws.amazon.com/braket/}.

\bibitem{google_quantum}
K.~Kissell and N.~DeSantis, ``Expanding access to quantum today for a better
  tomorrow,''
  \url{https://cloud.google.com/blog/products/compute/ionq-quantum-computer-available-through-google-cloud}.

\bibitem{ibm_quantum}
``{Highlights of the IBM Quantum Summit 2022, IBM},''
  \url{https://www.ibm.com/quantum}.

\bibitem{azure_quantum}
``{Azure Quantum, Microsoft},''
  \url{https://azure.microsoft.com/en-us/products/quantum/}.

\bibitem{ccgrid2023showcase}
R.~Sangle, T.~Khare, P.~Seshadri, and Y.~Simmhan, ``Comparing the orchestration
  of quantum applications on hybrid clouds,'' in \emph{Students' Showcase, The
  23rd IEEE/ACM International Symposium on Cluster, Cloud and Internet
  Computing (CCGRID)}, 2023.

\bibitem{cerezo2021variational}
M.~Cerezo, A.~Arrasmith, R.~Babbush, S.~C. Benjamin, S.~Endo, K.~Fujii, J.~R.
  McClean, K.~Mitarai, X.~Yuan, L.~Cincio \emph{et~al.}, ``Variational quantum
  algorithms,'' \emph{Nature Reviews Physics}, vol.~3, no.~9, pp. 625--644,
  2021.

\bibitem{preskill2018quantum}
J.~Preskill, ``Quantum computing in the nisq era and beyond,'' \emph{Quantum},
  vol.~2, p.~79, 2018.

\bibitem{peng2020simulating}
\BIBentryALTinterwordspacing
T.~Peng, A.~W. Harrow, M.~Ozols, and X.~Wu, ``Simulating large quantum circuits
  on a small quantum computer,'' \emph{Physical Review Letters}, vol. 125,
  no.~15, p. 150504, 2020. [Online]. Available:
  \url{https://doi.org/10.1103/PhysRevLett.125.150504}
\BIBentrySTDinterwordspacing

\bibitem{adhikari2019survey}
M.~Adhikari, T.~Amgoth, and S.~N. Srirama, ``A survey on scheduling strategies
  for workflows in cloud environment and emerging trends,'' \emph{ACM Computing
  Surveys (CSUR)}, vol.~52, no.~4, pp. 1--36, 2019.

\bibitem{topcuoglu2002performance}
H.~Topcuoglu, S.~Hariri, and M.-Y. Wu, ``Performance-effective and
  low-complexity task scheduling for heterogeneous computing,'' \emph{IEEE
  transactions on parallel and distributed systems}, vol.~13, no.~3, pp.
  260--274, 2002.

\bibitem{yu2005cost}
J.~Yu, R.~Buyya, and C.~K. Tham, ``Cost-based scheduling of scientific workflow
  applications on utility grids,'' in \emph{First International Conference on
  e-Science and Grid Computing (e-Science'05)}.\hskip 1em plus 0.5em minus
  0.4em\relax Ieee, 2005, pp. 8--pp.

\bibitem{bousselmi2016energy}
K.~Bousselmi, Z.~Brahmi, and M.~M. Gammoudi, ``Energy efficient partitioning
  and scheduling approach for scientific workflows in the cloud,'' in
  \emph{2016 IEEE International Conference on Services Computing (SCC)}.\hskip
  1em plus 0.5em minus 0.4em\relax IEEE, 2016, pp. 146--154.

\bibitem{ravi2021}
G.~S. Ravi, K.~N. Smith, P.~Murali, and F.~T. Chong, ``Adaptive job and
  resource management for the growing quantum cloud,'' in \emph{2021 IEEE
  International Conference on Quantum Computing and Engineering (QCE)}.\hskip
  1em plus 0.5em minus 0.4em\relax IEEE, 2021, pp. 301--312.

\bibitem{zhang2022}
M.~Zhang, Y.~Fu, J.~Wang, and J.~Lai, ``Research on task scheduling scheme for
  quantum computing cloud platform,'' in \emph{Proceedings of the 2022 6th
  International Conference on Cloud and Big Data Computing}, 2022, pp. 7--11.

\bibitem{parekh2021}
R.~Parekh, A.~Ricciardi, A.~Darwish, and S.~DiAdamo, ``Quantum algorithms and
  simulation for parallel and distributed quantum computing,'' in \emph{2021
  IEEE/ACM Second International Workshop on Quantum Computing Software
  (QCS)}.\hskip 1em plus 0.5em minus 0.4em\relax IEEE, 2021, pp. 9--19.

\bibitem{perlin2021quantum}
\BIBentryALTinterwordspacing
M.~A. Perlin, Z.~H. Saleem, M.~Suchara, and J.~C. Osborn, ``Quantum circuit
  cutting with maximum-likelihood tomography,'' \emph{npj Quantum Information},
  vol.~7, no.~1, pp. 1--8, 2021. [Online]. Available:
  \url{https://doi.org/10.48550/arXiv.2005.12702}
\BIBentrySTDinterwordspacing

\bibitem{tang2021cutqc}
\BIBentryALTinterwordspacing
W.~Tang, T.~Tomesh, M.~Suchara, J.~Larson, and M.~Martonosi, ``Cutqc: using
  small quantum computers for large quantum circuit evaluations,'' in
  \emph{Proceedings of the 26th ACM International Conference on Architectural
  Support for Programming Languages and Operating Systems}, 2021, pp. 473--486.
  [Online]. Available: \url{https://doi.org/10.1145/3445814.3446758}
\BIBentrySTDinterwordspacing

\bibitem{brenner2023optimal}
L.~Brenner, C.~Piveteau, and D.~Sutter, ``Optimal wire cutting with classical
  communication,'' \emph{arXiv preprint arXiv:2302.03366}, 2023.

\bibitem{piveteau2022circuit}
C.~Piveteau and D.~Sutter, ``Circuit knitting with classical communication,''
  \emph{arXiv preprint arXiv:2205.00016}, 2022.

\bibitem{marshall2022}
S.~C. Marshall, C.~Gyurik, and V.~Dunjko, ``High dimensional quantum learning
  with small quantum computers,'' \emph{arXiv preprint arXiv:2203.13739}, 2022.

\bibitem{eqc}
S.~Stein, N.~Wiebe, Y.~Ding, P.~Bo, K.~Kowalski, N.~Baker, J.~Ang, and A.~Li,
  ``Eqc: ensembled quantum computing for variational quantum algorithms,'' in
  \emph{Proceedings of the 49th Annual International Symposium on Computer
  Architecture}, 2022, pp. 59--71.

\bibitem{QiskitRuntime}
``{Qiskit Runtime, IBM},'' \url{https://www.ibm.com/quantum/qiskit-runtime}.

\bibitem{viola1998dynamical}
L.~Viola and S.~Lloyd, ``Dynamical suppression of decoherence in two-state
  quantum systems,'' \emph{Physical Review A}, vol.~58, no.~4, p. 2733, 1998.

\bibitem{van2022}
E.~Van Den~Berg, Z.~K. Minev, and K.~Temme, ``Model-free readout-error
  mitigation for quantum expectation values,'' \emph{Physical Review A}, vol.
  105, no.~3, p. 032620, 2022.

\bibitem{temme2017}
K.~Temme, S.~Bravyi, and J.~M. Gambetta, ``Error mitigation for short-depth
  quantum circuits,'' \emph{Physical review letters}, vol. 119, no.~18, p.
  180509, 2017.

\bibitem{van2022probabilistic}
E.~van~den Berg, Z.~K. Minev, A.~Kandala, and K.~Temme, ``Probabilistic error
  cancellation with sparse pauli-lindblad models on noisy quantum processors,''
  \emph{arXiv e-prints}, pp. arXiv--2201, 2022.

\bibitem{peruzzo2014variational}
A.~Peruzzo, J.~McClean, P.~Shadbolt, M.-H. Yung, X.-Q. Zhou, P.~J. Love,
  A.~Aspuru-Guzik, and J.~L. O’brien, ``A variational eigenvalue solver on a
  photonic quantum processor,'' \emph{Nature communications}, vol.~5, no.~1, p.
  4213, 2014.

\bibitem{QSVM}
V.~Havlíček, A.~D. Córcoles, K.~Temme, A.~W. Harrow, A.~Kandala, J.~M. Chow,
  and J.~M. Gambetta, ``Supervised learning with quantum-enhanced feature
  spaces.'' \emph{Nature}, pp. 209--212, 2019.

\bibitem{basu2021qer}
\BIBentryALTinterwordspacing
S.~Basu, A.~Saha, A.~Chakrabarti, and S.~Sur-Kolay, ``i-qer: An intelligent
  approach towards quantum error reduction,'' \emph{ACM Transactions on Quantum
  Computing}, 2021. [Online]. Available:
  \url{https://dl.acm.org/doi/10.1145/3539613}
\BIBentrySTDinterwordspacing

\bibitem{ayral2021quantum}
\BIBentryALTinterwordspacing
T.~Ayral, F.-M.~L. R{\'e}gent, Z.~Saleem, Y.~Alexeev, and M.~Suchara, ``Quantum
  divide and compute: exploring the effect of different noise sources,''
  \emph{SN Computer Science}, vol.~2, no.~3, pp. 1--14, 2021. [Online].
  Available: \url{https://doi.org/10.1007/s42979-021-00508-9}
\BIBentrySTDinterwordspacing

\bibitem{majumdar2022error}
R.~Majumdar and C.~J. Wood, ``Error mitigated quantum circuit cutting,''
  \emph{arXiv preprint arXiv:2211.13431}, 2022.

\bibitem{saleem2021quantum}
Z.~H. Saleem, T.~Tomesh, M.~A. Perlin, P.~Gokhale, and M.~Suchara, ``Quantum
  divide and conquer for combinatorial optimization and distributed
  computing,'' \emph{arXiv preprint arXiv:2107.07532}, 2021.

\bibitem{smith2023clifford}
K.~N. Smith, M.~A. Perlin, P.~Gokhale, P.~Frederick, D.~Owusu-Antwi, R.~Rines,
  V.~Omole, and F.~T. Chong, ``Clifford-based circuit cutting for quantum
  simulation,'' \emph{arXiv preprint arXiv:2303.10788}, 2023.

\bibitem{circuit_knitting}
``\text{Circuit Knitting Toolbox},''
  \url{https://qiskit-extensions.github.io/circuit-knitting-toolbox/index.html}.

\bibitem{HFdatapaper}
D.~Chicco and G.~Jurman, ``Machine learning can predict survival of patients
  with heart failure from serum creatinine and ejection fraction alone,''
  \emph{BMC Medical Informatics and Decision Making}, vol.~20, 2020.

\bibitem{wack2021scale}
A.~Wack, H.~Paik, A.~Javadi-Abhari, P.~Jurcevic, I.~Faro, J.~M. Gambetta, and
  B.~R. Johnson, ``Scale, quality, and speed: three key attributes to measure
  the performance of near-term quantum computers,'' \emph{arXiv preprint
  arXiv:2110.14108}, 2021.

\end{thebibliography}

\end{document}